\documentclass[journal=jacsat,manuscript=article,layout=twocolumn]{achemso}

\usepackage{geometry}

\setkeys{acs}{articletitle=true}
\usepackage{lettrine}

\usepackage[version=3]{mhchem} % Formula subscripts using \ce{}
\usepackage[T1]{fontenc}       % Use modern font encodings

\usepackage[fontsize=10pt]{scrextend}

\usepackage{newtxtext}
\usepackage{bm} 

\usepackage{titlesec} % inline subsections
\titleformat{\subsection}[runin]
{\normalfont\bfseries}{\thesubsection{.}}{1em}{}[.]

\titleformat{\section}
{\normalfont\sffamily\large\bfseries}{\thesection{.}}{1em}{\MakeUppercase}

\usepackage[table]{xcolor}
\usepackage{graphicx}
\usepackage{amsmath,fullpage,amssymb}
\usepackage{graphics,epsfig}
\usepackage{makeidx}
\usepackage{gensymb}
\usepackage{siunitx}

\newcommand*{\citen}[1]{% in-line citation
  \begingroup
    \romannumeral-`\x % remove space at the beginning of \setcitestyle
    \setcitestyle{numbers}%
    \cite{#1}%
  \endgroup   
}

\def\rmd{{\mathrm{d}}}

\def\rme{{\mathrm{e}}}

\def\Eq{eq}
\def\Eqs{eqs}
\def\Figure{\textcolor{black}{Figure}}
\def\Fig{\textcolor{black}{Figure}}

\def\SItext{\textcolor{black}{Supporting Information}}
\def\ie{{\em i.e.}}
\def\eg{{\em e.g.}}
\def\via{{\em via}}

\newcommand{\Av}[1]{{\bf #1}}

\newcommand{\kB}{k_\textrm{B}}
\newcommand{\chgA}[1]{{#1}}

\newcommand{\trm}[1]{{\textrm{#1}}}

\usepackage{soul}	% for strikethrough font

\SectionNumbersOn
\let\oldmaketitle\maketitle
\let\maketitle\relax

\title{\flushleft Aqueous Nanoclusters Govern Ion Partitioning in Dense Polymer Membranes
}

\author{Matej Kandu\v{c}}
\affiliation{\rm\small Jo\v{z}ef Stefan Institute, Jamova 39, 1000 Ljubljana, Slovenia}
\email{matej.kanduc@ijs.si}

\author{Won Kyu Kim}
\affiliation{\rm\small Korea Institute for Advanced Study, 85 Hoegiro, Seoul 02455, Republic of Korea}
\alsoaffiliation{\rm\small Freie Universit\"{a}t Berlin, Fachbereich Physik, Arnimallee 14, 14195 Berlin, Germany}
\alsoaffiliation{\rm\small Research Group for Simulations of Energy Materials, Helmholtz-Zentrum Berlin f\"ur Materialien und Energie, Hahn-Meitner-Platz 1, 14109 Berlin, Germany}

\author{Rafael Roa}
\affiliation{\rm\small Departamento de F\'{i}sica Aplicada I, Facultad de Ciencias, Universidad de M\'{a}laga, Campus de Teatinos s/n, 29071 M\'{a}laga, Spain}

\author{Joachim Dzubiella}
\affiliation{\rm\small Applied Theoretical Physics -- Computational Physics, Physikalisches Institut, Albert-Ludwigs-Universit\"at Freiburg, Hermann-Herder Strasse 3, D-79104 Freiburg, Germany}
\alsoaffiliation{\rm\small Research Group for Simulations of Energy Materials, Helmholtz-Zentrum Berlin f\"ur Materialien und Energie, Hahn-Meitner-Platz 1, D-14109 Berlin, Germany}
\email{joachim.dzubiella@physik.uni-freiburg.de}

\begin{document}
\pagenumbering{arabic}
\noindent

\parindent=0cm
\setlength\arraycolsep{2pt}
\renewcommand{\abstractname}{\large \textsf{ABSTRACT}}
\renewcommand\refname{\large\textsf{REFERENCES}}

\twocolumn[	% make wide abstract
\begin{@twocolumnfalse}
\oldmaketitle

\begin{abstract}\small
The uptake and sorption of charged molecules by responsive polymer membranes and hydrogels in aqueous solutions is of key importance for the development of soft functional materials.  Here we investigate the partitioning of simple monoatomic (Na$^+$, K$^+$, Cs$^+$, Cl$^-$, I$^-$) and one molecular ion (4-nitrophenolate; NP$^-$) within a dense, electroneutral poly($N$-isopropylacrylamide) membrane using explicit-water molecular dynamics simulations. Inside the predominantly hydrophobic environment water distributes in a network of polydisperse water nanoclusters. The average cluster size determines the mean electrostatic self-energy of the simple ions, which preferably reside deeply inside them; we therefore find substantially larger partition ratios $K\simeq\>$10$^{-1}$ than expected from a simple Born picture using a uniform dielectric constant.  Despite their irregular shapes we observe that the water clusters possess a universal negative electrostatic potential with respect to their surrounding, as is known for aqueous liquid--vapor interfaces. This potential, which we find concealed in cases of symmetric monoatomic salts, can dramatically impact the transfer free energies of larger charged molecules because of their weak hydration and increased affinity to interfaces. Consequently, and in stark contrast to the simple ions,  the molecular ion NP$^-$ can have a partition ratio much larger than unity, $K\simeq\>$10--30 (depending on the cation type) or even \chgA{$10^3$ in excess of monovalent salt}, which explains recent observations of enhanced reaction kinetics of NP$^-$ reduction catalyzed within dense polymer networks. These results also suggest that ionizing a molecule can even enhance the partitioning in a collapsed, rather hydrophobic gel, which strongly challenges the traditional simplistic reasoning. 
\\
\\
{{\large\bf\textsf{KEYWORDS:}}} {hydrogel, hydration, ion solvation, surface potential, partitioning, molecular dynamics simulation}
\vspace{5ex}
\end{abstract}
\end{@twocolumnfalse}]

\maketitle
\setlength\arraycolsep{2pt}
\small

%Wijmans_Baker-JMemSci1995.pdf

The principal factor in the design of polymer membranes and hydrogels used in applications for water purification, pervaporation, drug delivery, nanocarriers, {\em etc.}\ is the capability to control the uptake and permeability of different molecular species~\cite{garba1999ion, wijmans1995solution,nykanen2007phase, chen2014thermo, hyk2018water, stuartNatMat2010, yingApplicationsSM2011, li2016designing, qiao2018modeling}.
  %Kamcev_Manning-PCCP2016.pdf
 For instance, the ability to control the permeability of ions makes such materials interesting
for water purification technologies, such as reverse and forward osmosis~\cite{kamcev2016partitioning, cath2006forward}.
  A separation is achieved between different penetrants because of distinction in the amount of material that dissolves in the membrane and the rate at which the material diffuses through the membrane~\cite{wijmans1995solution,rud2018thermodynamic}.  
  % Rud_Kosovan_Desalination2018.pdf --- compression
A recent modification of forward osmosis, where the hydrogel acts as the draw solute (osmotic agent) and the separation membrane  at the same time~\cite{hopfner2010novel, cai2016critical, arens2017energy}, exploits the difference of the salt concentration inside the hydrogel in its swollen and collapsed (or compressed state), and can be used to yield low-salinity water upon repeating the cycle~\cite{hopfner2010novel}. 

 As an alternative to mechanical compression, thermo-responsive shrinking of a hydrogel can be utilized for desalination purposes~\cite{li2011stimuli, ali2015design, rud2018thermodynamic}.
 Thermo-responsive polymers, such as poly($N$-isopropylacrylamide) (PNIPAM), undergo a transition from good to poor solvent conditions upon an increase in temperature or triggered by some other environmental stimulus, which results in substantial shrinkage of the gel~\cite{pelton2000, hudsonProgPolySci2004, halperin2015poly, schmid2016multi}. An advantage of responsive hydrogels in purification lies in simple separation by filtration methods and their reusability. Thus, polymer membranes are expected to continue to dominate the water purification technologies owing to their energy efficiency~\cite{geise2010water, rud2018thermodynamic}.

%is lower than outside [6,10]. In the experiments, is swollen in the solution of NaCl, and then compressed to yield water with a lower salt concentration than the initial solution [6]. The originally proposed desalination cycle included a recovery step by re-swelling the compressed gel in the high-salinity solution. This step is thermodynamically irreversible, which limits the maximum efficiency that can be theoretically achieved in this cycle [11]. The recent experimental examination of desalination effi- ciency of this cycle came to similar conclusions [7]. 
  %
%Improved fundamental understanding of ion sorption in polymers would offer insight into key factors governing ion transport through ion exchange polymers and as such accelerate development of high performance materials.

%Thus, it is important to establish a theory taking into account the membrane features and the operating conditions in order to predict the salt rejection.

% Li_Mooney-gels_drug_delivery-NatRev2016.pdf
Hydrogels are also particularly appealing materials for drug delivery systems 
as they can be designed in numerous ways in order to selectively encapsulate and release  particular types of pharmacologically relevant molecules in a controllable manner~\cite{li2016designing}.
For charged drugs, the electrostatic interactions, facilitated by accompanying ions,
 are an important, if not dominating, driving force for the transport through the gel~\cite{molinaPolymer2012}.
Related phenomena are exploited in adaptive nanocatalysis
where catalytic nanoparticles are confined in a permeable hydrogel that can be potentially employed as a programmable nanoreactor to shelter and control the catalytic process~\cite{jandtCatal2010, ballauffRoyal2009,carregal2010catalysis, dzubiellaAngew2012, roa2017catalyzed}.
Since most of the penetrating molecules in these applications, such as salt, ligands, and reactants are charged, the understanding and control of the behavior of simple and molecular ions inside hydrogels are critical for the development of high performance materials for the applications mentioned above. 

Two basic quantities that need first attention here are the solubility and the mobility (or diffusivity) of the molecules inside the polymer. The solubility is usually simply expressed by the partition ratio $K$, which describes the ratio between mean solute concentrations inside the membrane and in bulk~\cite{IUPACgoldbook}. The product of partitioning and diffusivity defines the membrane permeability  within the linear response solution--diffusion model~\cite{wijmans1995solution}.  In fact, because of a lower water content, a collapsed hydrogel exhibits an apparently more hydrophobic environment than in the swollen state, therefore the partitioning of ions is typically below unity (that is, below bulk concentration)~\cite{peng2012ion}. In a simple Born solvation picture, one could argue that the mean dielectric constant in a dense hydrogel is one order of magnitude lower than in bulk water and thus the transfer from bulk is strongly penalized by a lack of polarization. This penalty (see \Eq~\ref{eq:simple} later) would lead to partition ratios orders of magnitude smaller than observed for simple salts such as NaCl~\cite{peng2012ion}, a fact that is essentially unexplained. More strikingly, more complex, molecular ions  exhibit a much larger span of partitioning values, in some cases (\eg, small ionized drugs) even exceeding unity by orders of magnitude, $K\simeq10^3$,  in collapsed hydrogels~\cite{guilherme2003hydrogels, molinaPolymer2012}. This clearly eludes any simple electrostatic and mean-field mechanisms. Hence, despite a rich history of research on synthetic gels and ion exchange polymers, the present knowledge does still not suffice for quantitative, predictive connections between polymer structure and the thermodynamics of ion solvation in dense membranes~\cite{garba1999ion, geise2010water,kamcev2016partitioning}.

The reason for the lack of understanding must be sought in the complex interplay between water, ions, and the polymer matrix on the molecular level, particularly in dense polymers and collapsed hydrogels, where a very crowded and heterogeneous molecular environment can be expected~\cite{merrill1993partitioning}.
In fact, it has been shown by us recently using molecular simulations that the water--polymer spatial heterogeneity in a collapsed PNIPAM polymer affects the distribution of neutral solutes~\cite{kanduc2019free}, therefore one may anticipate something similar for ions. We studied previously also the diffusion of ions in the same dense polymer matrix where magnitude and scaling of diffusion with ionic size were very different from nonpolar solutes~\cite{kanduc2018diffusion}, which could be clearly traced back to local hydration effects. When comparing the interaction and adsorption of simple monoatomic ions and molecular ions~\cite{kanduc2017selective} to a single PNIPAM chain we observed that the former were repelled from the chain and stayed nicely hydrated in bulk, while the latter could easily adsorb to the chain. Hence, ionic solvation in a dense, weakly hydrated polymer matrix is not easily accessible and is challenging to categorize without a deeper molecular level understanding. 

In this study, we aim for such a detailed understanding of the problem of ion partitioning in the dense, weakly hydrated polymers by using classical molecular dynamics simulations. As a case study we choose the popular PNIPAM polymer in the collapsed state, above the transition temperature, where the water amount is low (20 wt\%). We perform a detailed analysis of ionic solvation free energies in bulk water and in the membrane as well as of hydration structure in the polymer matrix.  We discuss in detail the electrostatic contributions to the salt and ionic partition ratios and devise a simple solvation theory that explains the simulation outcomes. At the end we devote special attention to the comparison of simple monoatomic ions to the molecular ion 4-nitrophenolate (NP$^-$), which behaves markedly different because of weaker hydration and its affinity to the heterogeneous nanosized interfaces within the membrane. This study thus provides  molecular insight into ion partitioning in a dense polymer and will be helpful for the interpretation of various experimental data, directly or indirectly related to ionic concentrations inside the polymeric functional material. 

\section*{Results and discussions}

% Geise_Freeman_ProgPolySci2014.pdf
\subsection*{Background}
The uptake of salt ions and larger charged (drug-like) molecules by a hydrogel depends crucially on the molecular details (the ``chemistry'') of the polymer matrix. It is characterized by the partition ratio $K$~\cite{IUPACgoldbook}, and is for low enough concentrations independent of the concentration.

\begin{figure}[h]\begin{center}
\begin{minipage}[b]{0.38\textwidth}\begin{center}
\includegraphics[width=\textwidth]{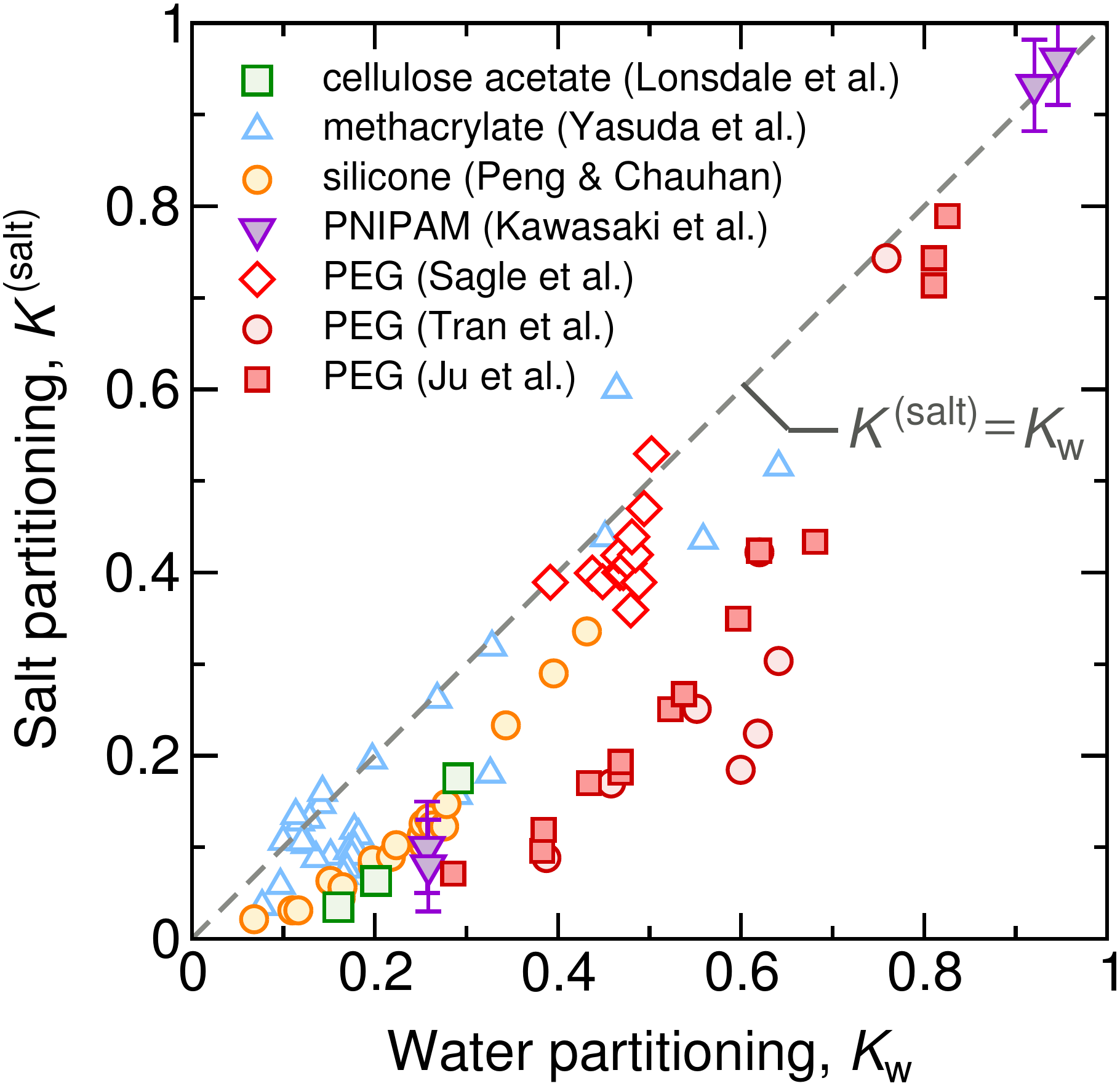}
\end{center}\end{minipage}
\caption{Correlation between NaCl partition ratio, $K^\trm{(salt)}$, and the water partition ratio, $K_\trm{w}$, for several polymers reported in the literature: methacrylate~\cite{yasuda1968permeability}, cellulose acetate~\cite{lonsdale1965transport}, silicone~\cite{peng2012ion}, PNIPAM~\cite{kawasaki2000partition}, and PEG~\cite{sagle2009peg,tran2019elucidating, ju2010characterization} at different temperatures or with different degrees of copolymerization.
%Series of hydrogel materials ... containing various fractions of copolymers and crosslinkers
%Hydrogels based on (various fractions of copolymers and crosslinkers) with primary polymers based on methacrylate~\cite{yasuda1968permeability}, cellulose acetate~\cite{lonsdale1965transport}, silicone~\cite{peng2012ion}, and PEG~\cite{sagle2009peg,tran2019elucidating, ju2010characterization}.
}
\label{fig:K}
\end{center}\end{figure}

In \Fig~\ref{fig:K} we show the correlation between the NaCl salt ($K^\trm{(salt)}$) and water ($K_\trm{w}$) partition ratios (the latter defined as the ratio of water densities inside and outside the gel)
for several uncharged polymers reported in the literature. Here, the primary polymer [methacrylate~\cite{yasuda1968permeability}, cellulose acetate~\cite{lonsdale1965transport}, silicone~\cite{peng2012ion}, PNIPAM~\cite{kawasaki2000partition}, or polyethylene glycol (PEG)~\cite{sagle2009peg,tran2019elucidating, ju2010characterization}] was cross-linked and either exposed to different temperatures or copolymerized with other monomers   in order to tune the equilibrium water content.
Quite generally, the uptake of salt by a polymer depends to some extent on water amount; polymers that sorb more water often tend to sorb more salt than those polymers that sorb less water~\cite{geise2014fundamental}. 
Also, most of the data fall below the dashed line, depicting an apparent limiting case of $K^\trm{(salt)} = K_\trm{w}$ where the salt concentration in the sorbed water is equal to that in bulk water~\cite{yasuda1968permeability}. But since mostly $K^\trm{(salt)} < K_\trm{w}$, this indicates that both polymer--ion and polymer--water interactions influence salt partitioning.

On the other hand, larger charged molecules (\eg, ionized drugs) exhibit a much larger span of partitioning values, in some cases even exceeding $K=10^3$ in collapsed hydrogels~\cite{guilherme2003hydrogels, molinaPolymer2012}. This clearly eludes the empirical observations for monoatomic salts and calls for additional mechanisms, which we will address in this study.

\begin{figure*}[t]\begin{center}
\begin{minipage}[b]{0.30\textwidth}\begin{center}
\includegraphics[width=\textwidth]{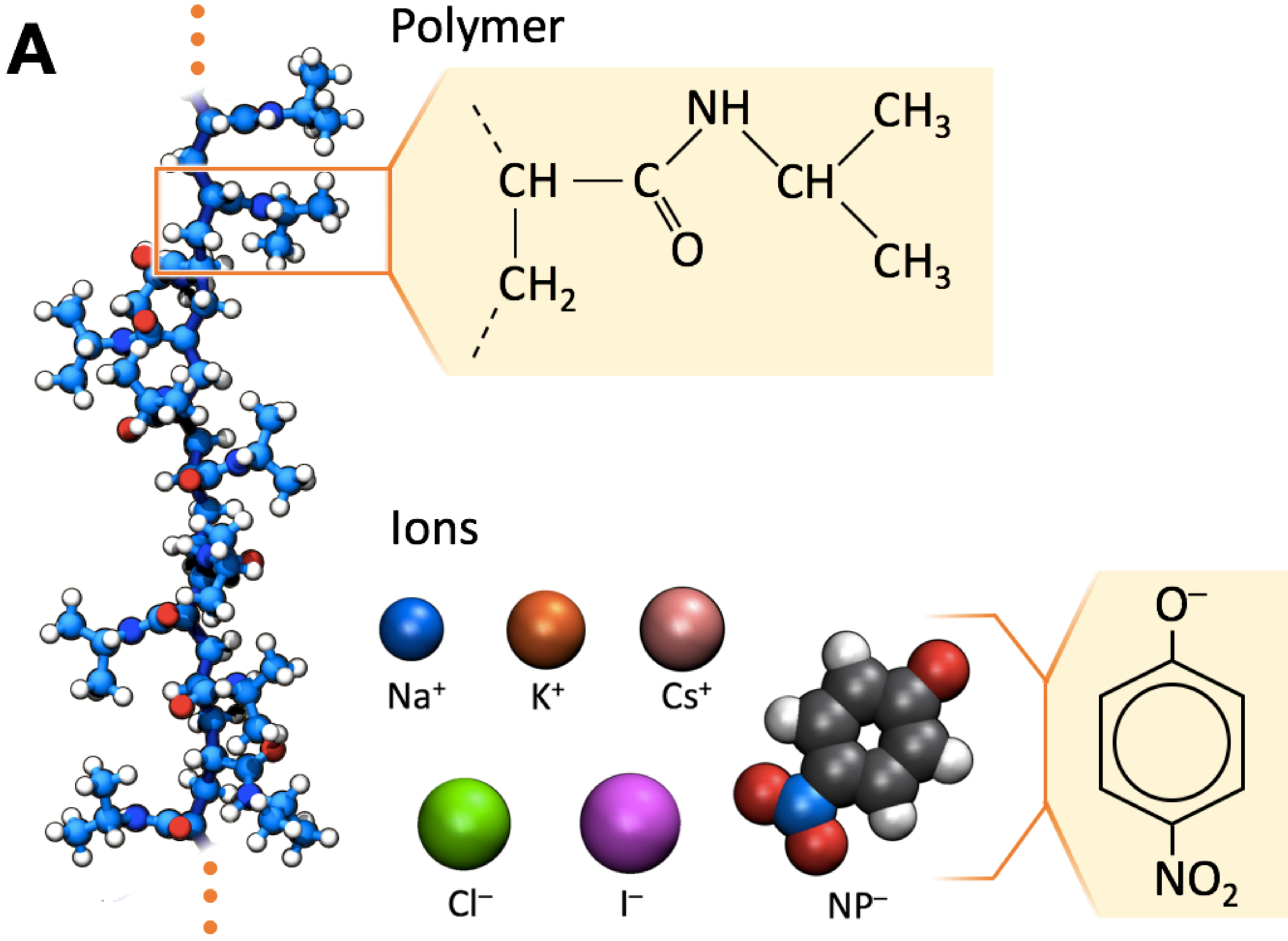}
\end{center}\end{minipage}\hspace{2ex}
\begin{minipage}[b]{0.254\textwidth}\begin{center}
\includegraphics[width=\textwidth]{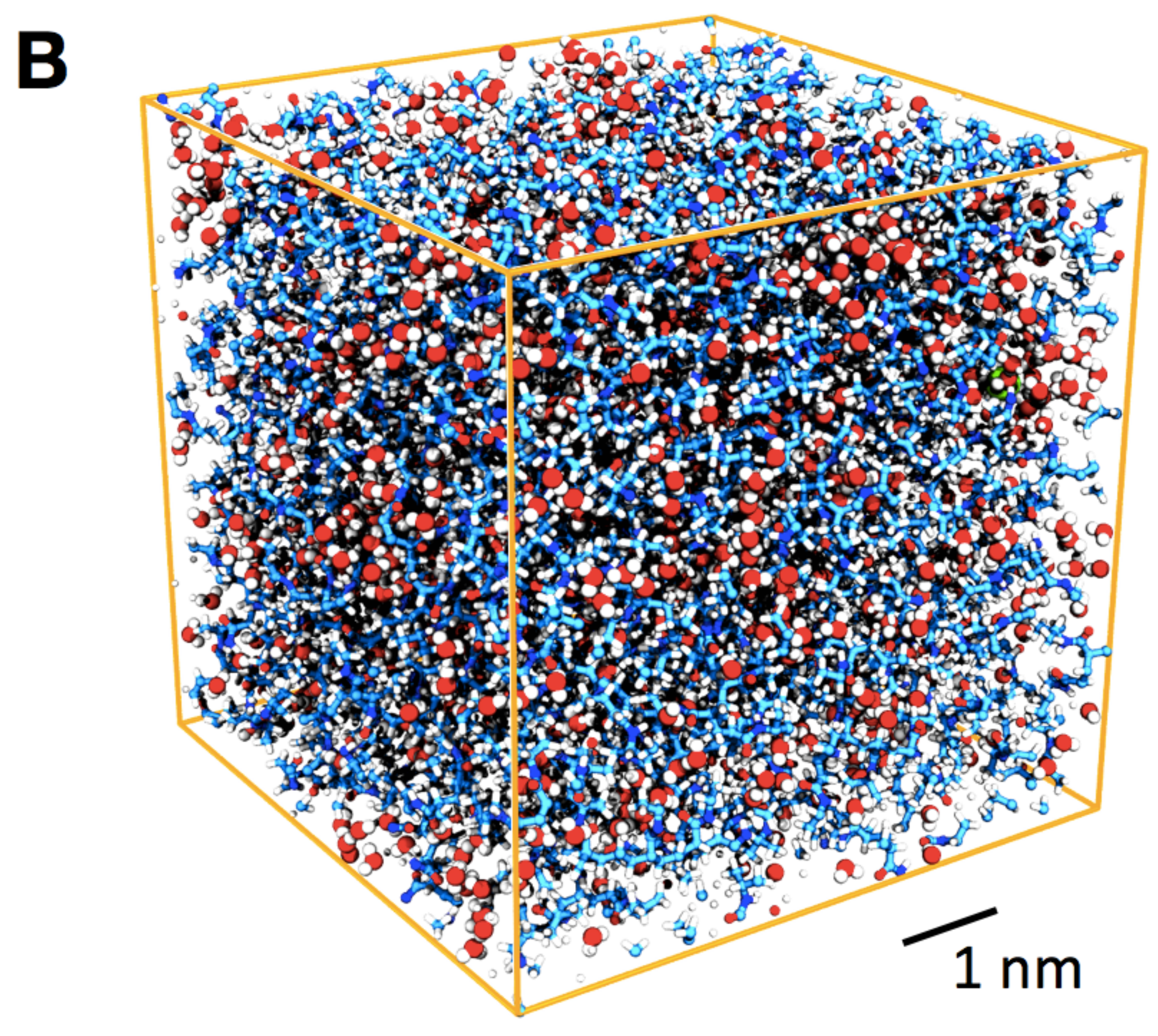}
\end{center}\end{minipage}\hspace{2.8ex}
\begin{minipage}[b]{0.34\textwidth}\begin{center}
\includegraphics[width=\textwidth]{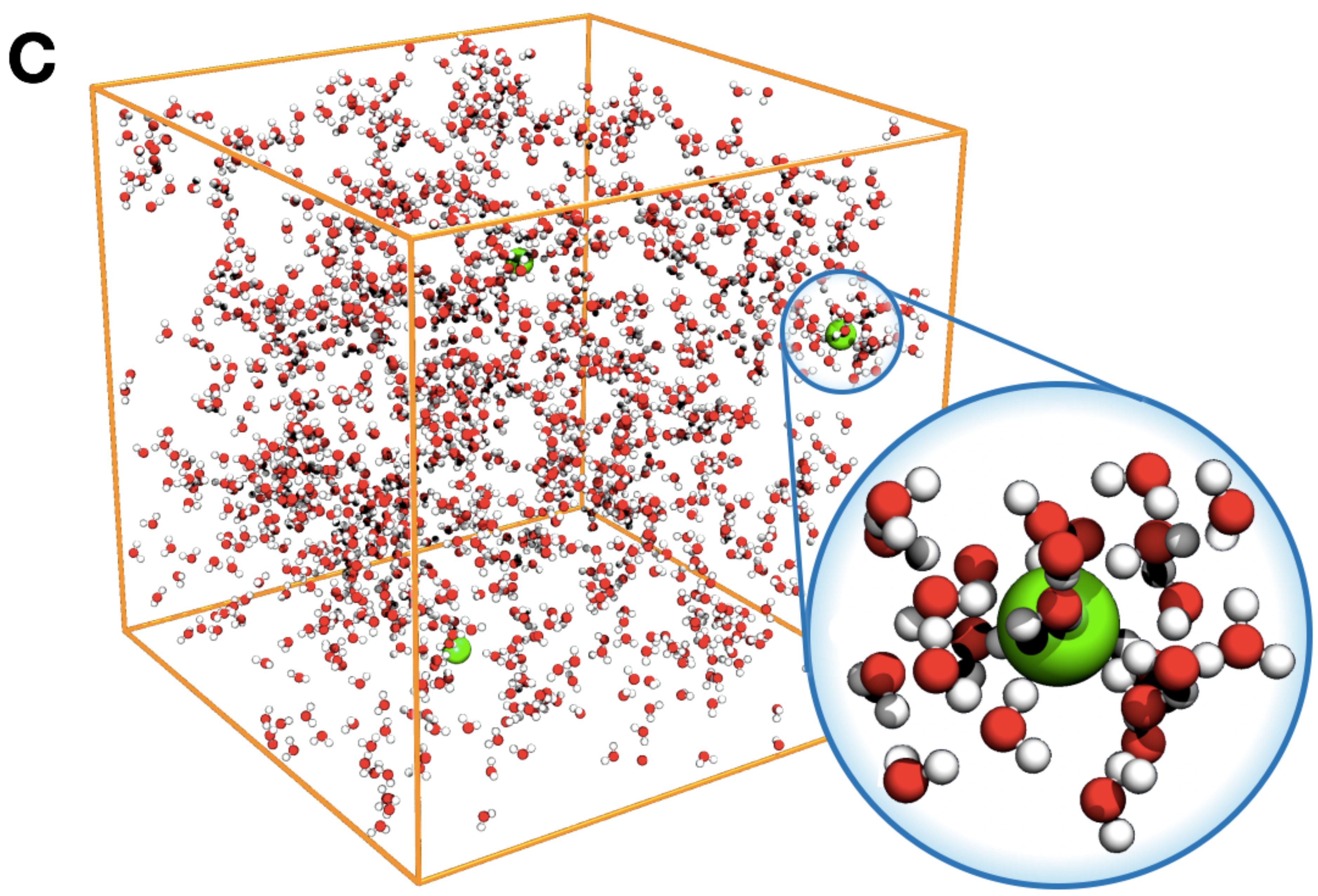}
\end{center}\end{minipage}
\caption{(A) PNIPAM polymer chain (in a stretched conformation) and ions in this study.
Snapshots of condensed polymers and water containing three chloride ions, showing (B)~all components  and (C)~water and ions only. The blue bubble zooms into the proximity of one of the Cl$^-$ ions, featuring its hydration shell.}
\label{fig:snap}
\end{center}\end{figure*}

\subsection*{Transfer free energies}
In this computer simulation study, we use a model of a collapsed PNIPAM polymer phase above the transition temperature with the water partition ratio $K_\trm{w}=$\,0.2 (see \Fig~\ref{fig:snap}A for a single extended polymer chain). This is also a good model of a collapsed hydrogel with very low (few percents) cross-linker concentrations.
% reword more:
A typical simulation snapshot is shown in \Fig~\ref{fig:snap}B with 48 polymer chains in blue and 1325 water molecules in red--white. Removing the polymer component from the plot (\Fig~\ref{fig:snap}C) reveals that water molecules are very non-uniformly distributed, forming irregular, lacy-like clusters, which engender a heterogeneous polar--nonpolar environment.
\chgA{Such water clustering and aggregation are otherwise well known to occur in various amorphous polymer structures~\cite{tamai1994molecular, fukuda1998clustering, kucukpinar2003molecular, goudeau2004atomistic,marque2008molecular}.}
%The water clusters are very polydisperse and, with approximately power-law distribution $P(N_\trm{w})\sim  N_\trm{w}^{-1.74}$, where $N_\trm{w}$ is the number of water molecules in the cluster.
The radius of gyration $R_\trm{g}$ of the clusters in our system scales with the number of water molecules $N_\trm{w}$ as $ R_\trm{g}\sim {N_\trm{w}}^{\!1/2}$ and hence retains some characteristic of the random walk.
More details about the structure can be found elsewhere~\cite{kanduc2018diffusion}.

%In this study we focus on the thermodynamic properties of monovalent ions inside this dense polymeric hydrogel. 
For simplicity, we restrict ourselves to the solvation of monovalent ions: three alkali metal cations (sodium, potassium, cesium), two halides (chloride and iodide), and a molecular ion 4-nitrophenolate (NP$^-$), which is 4-nitrophenol (NP$^0$) with a deprotonated hydroxyl group, see \Fig~\ref{fig:snap}A.
Nitrophenol became popular in model reactions in nanocatalytic benchmark experiments and was also used in a connection with PNIPAM hydrogels~\cite{dzubiellaAngew2012}.

In order to analyze the ionic partitioning we resort to free energy calculations (see Methods section). Note that a direct evaluation of $K$ in our system is not tractable due to a very slow diffusion of ions (see Methods section).
\begin{figure}[h]\begin{center}
\begin{minipage}[b]{0.37\textwidth}\begin{center}
\includegraphics[width=\textwidth]{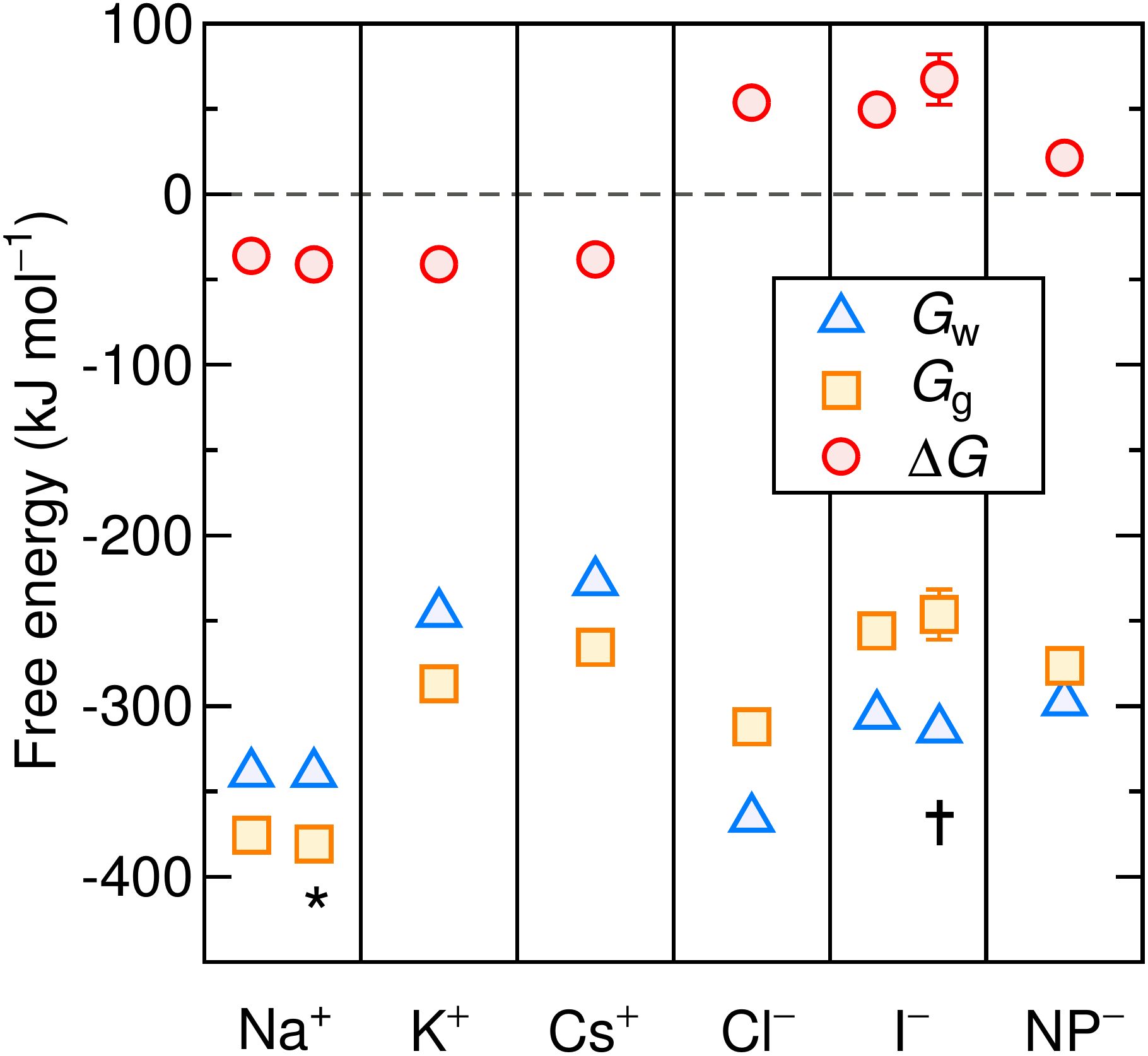}
\end{center}\end{minipage}
\caption{Free energies of ions: solvation free energy in water ($G_\trm{w}$), solvation free energy in PNIPAM gel ($G_\trm{g}$), and the transfer free energy from water into the gel ($\Delta G=G_\trm{g}-G_\trm{w}$).
For all monoatomic ions we use the Jorgensen force field~\cite{jorgensenFFNaCl, jorgensenFFI, jensen2006halide} and additionally $^\star$\AA qvist~\cite{aqvist1990ion} for Na$^+$ and $^{\dagger}$Dang~\cite{dang1992nonadditive,rajamani2004size} for I$^-$.
}
\label{fig:ions}
\end{center}\end{figure}
Figure~\ref{fig:ions} shows the evaluated solvation free energies of single ions in water ($G_\trm{w}$, blue triangle symbols) and in the gel ($G_\trm{g}$, orange square symbols).
They are negative and comparable in size (of the order of several $-$100~kJ/mol). 
Their difference corresponds to the central quantity of  interest in this study, namely to the transfer free energy from water into the gel, $\Delta G=G_\trm{g}-G_\trm{w}$.
Note that $\Delta G$ represents the {\sl real} single-ion solvation free energy because it is the one encompassing the totality of the reversible work associated with the physical process of transferring the ion from water into the gel.
In general, the experimental determination of single-ion solvation free energies is complicated 
because of the electroneutrality of macroscopic matter~\cite{hunenberger2011single}. Note that the gel--water interface potential is not included in $\Delta G$ because it is screened by ions sufficiently away from the interface compared with the Debye screening length (see \SItext).

The computed values of $\Delta G$ are depicted by red circles in \Fig~\ref{fig:ions}.
Evidently, we are confronting a very surprising and important observation: there is a stark contrast between the cations and anions in terms of their transfer free energies, $\Delta G$!
Systematically, all the monoatomic anions span in a narrow window of 50--70~kJ/mol, and are thus intrinsically repelled from the gel, whereas cations cover the negative range between $-$30 and $-$40~kJ/mol and should be attracted by the gel.
This difference stretches far beyond ion-specific effects (interpreted as the variation among different ions and force fields of the same valency).
Note that NP$^-$, as a molecular ion, includes additional contributions and will be discussed separately. 

This is surprising since in the traditional view one expects that charged entities should be universally repelled from a less polarizable medium. Furthermore, also atomistic simulation studies of isolated PNIPAM polymers and swollen networks utilizing similar or even the same force fields~\cite{du2013specificity, kanduc2017selective, milster2019crosslinker} show a universal repulsion of small cations and anions from the chains. 
Somehow, the aggregation of the polymers and expulsion of water (from excess of water down to 20 wt\% in our case) inverts the scenario for cations but not for anions.
Because the effect is triggered by reducing the water amount, this also suggests that the cation--anion asymmetry does most probably not stem from polymer--ion interactions, but has rather something to do with the water structure and its amount.

% surface polarization rezults into cluster potential

\subsection*{Monoatomic ions}
\begin{figure*}[t]\begin{center}
\begin{minipage}[b]{0.39\textwidth}\begin{center}
\includegraphics[width=\textwidth]{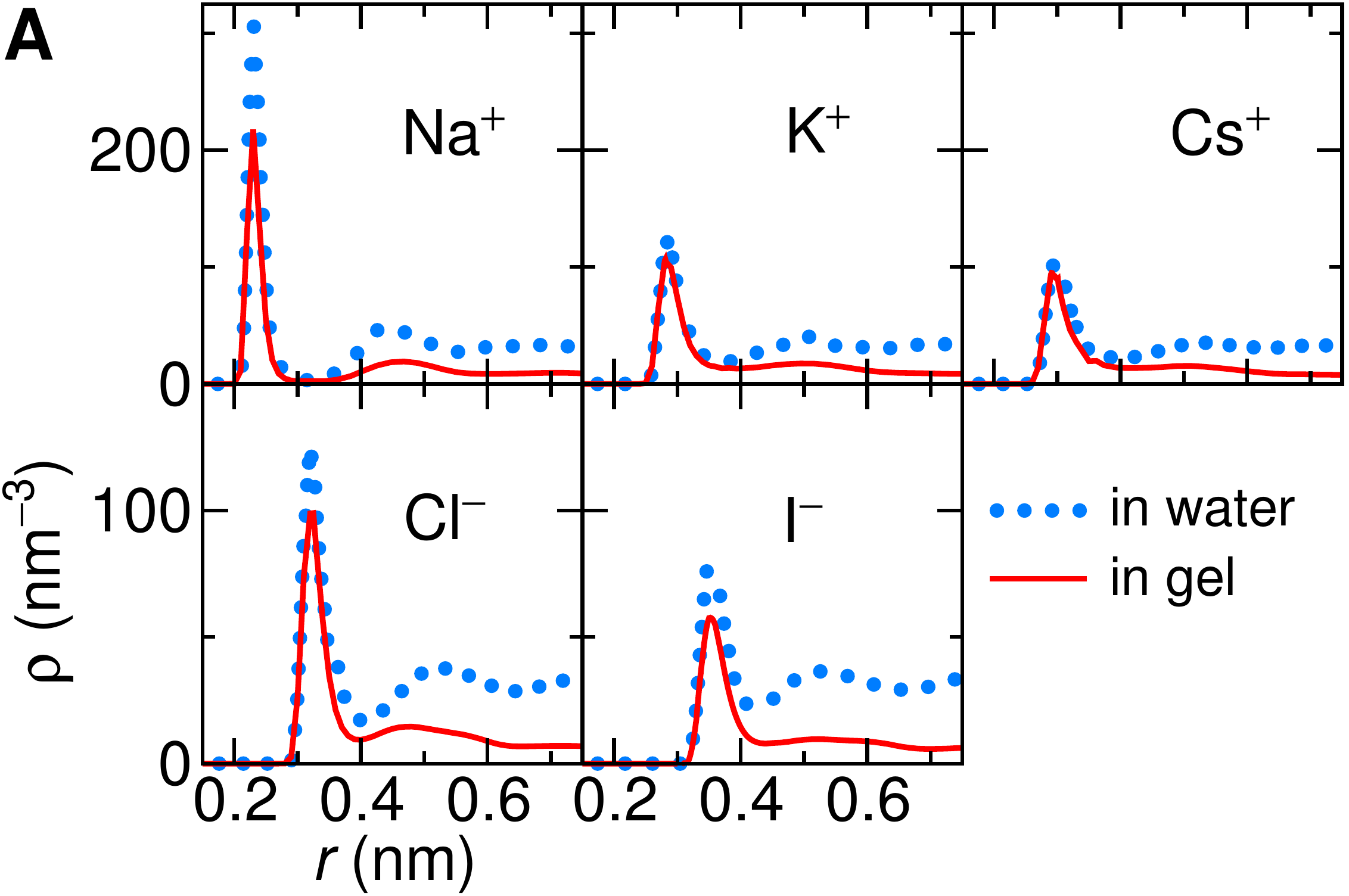} 
\end{center}\end{minipage}\hspace{2ex}
\begin{minipage}[b]{0.29\textwidth}\begin{center}
\includegraphics[width=\textwidth]{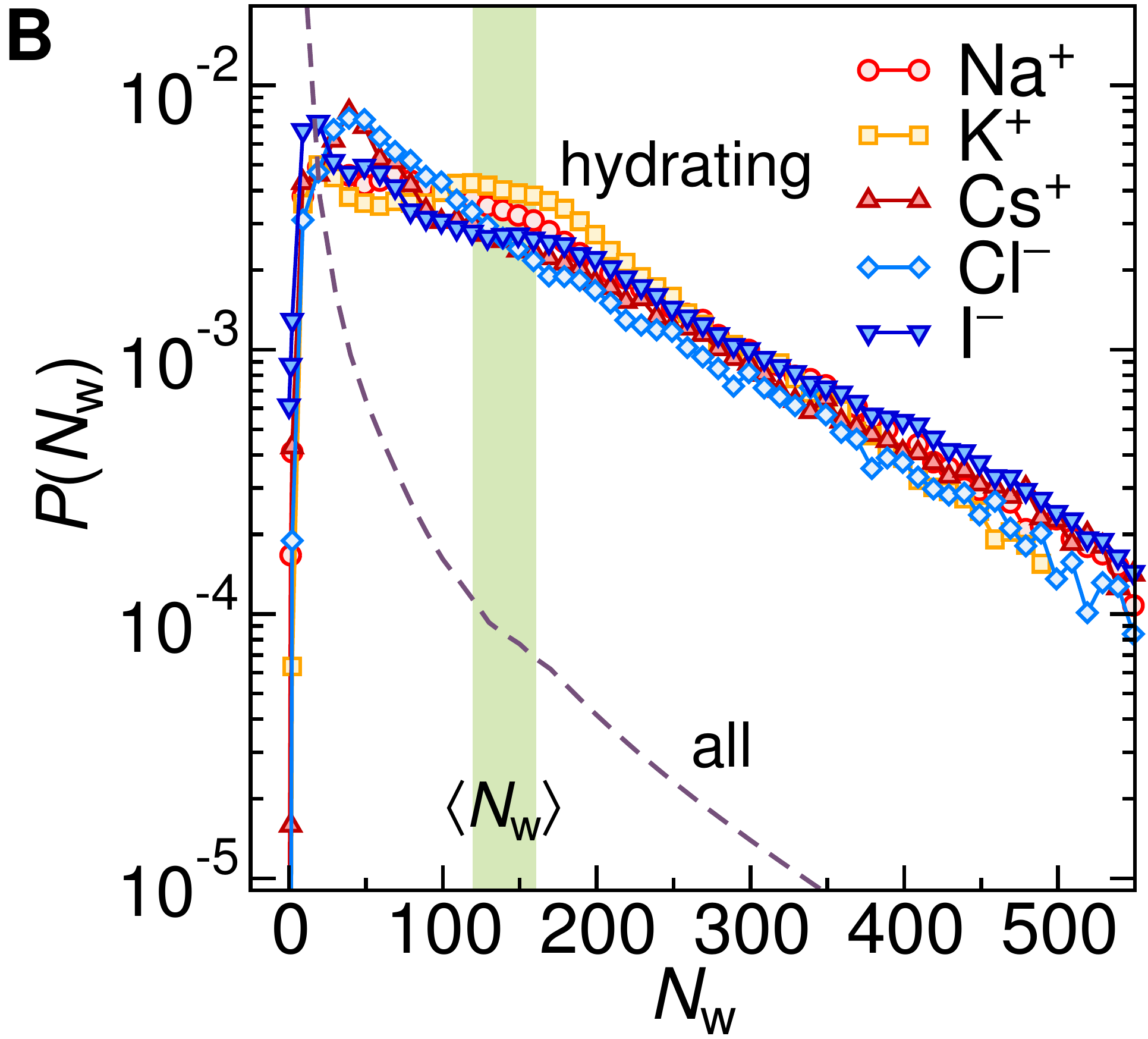}
\end{center}\end{minipage}\hspace{3ex}
\begin{minipage}[b]{0.237\textwidth}\begin{center}
\includegraphics[width=\textwidth]{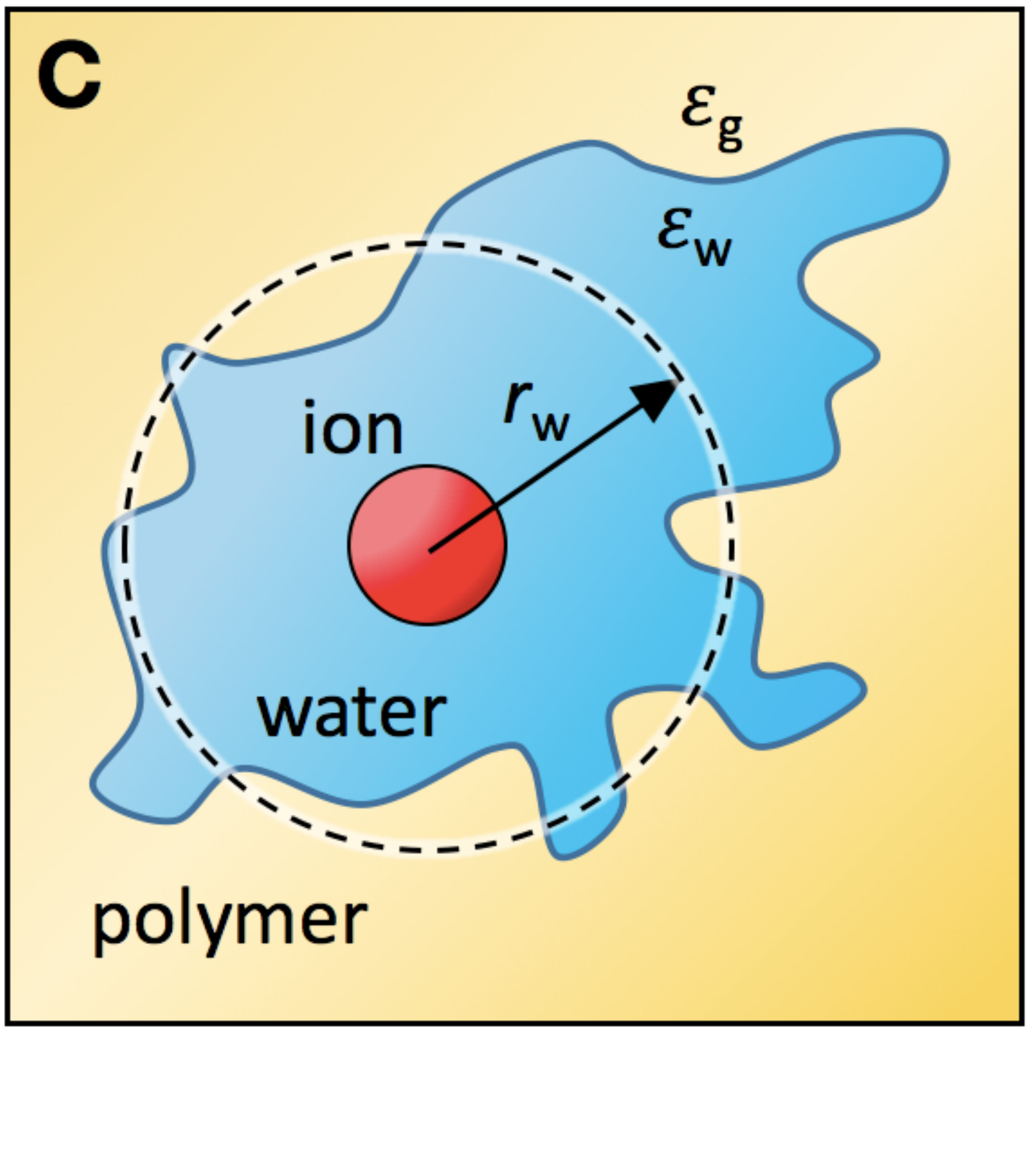}
\end{center}\end{minipage}
\caption{(A) Water density profiles around ions in bulk water (dotted lines) and PNIPAM (solid lines).
(B) Ion-hydrating cluster size distribution for different ions (symbols) and size distribution of all water clusters in the gel (dashed line). The green-shaded region indicates the mean ion-hydrating cluster size.
(C) Continuum picture of an ion encapsulated by a water cluster of an effective radius $r_\trm{w}$ embedded into a hydrogel environment.
}
\label{fig:RDFions}
\end{center}\end{figure*}

We will first focus on the mechanisms operative in the solvation of monoatomic ions.
For that, we begin by taking a closer look into the structural properties of the solvation.
It has already been shown that the water--polymer spatial heterogeneity in this collapsed polymer affects the distribution of neutral solutes~\cite{kanduc2019free}, therefore one can expect a related behavior for ions.
Indeed, already glancing at the representative snapshot in \Fig~\ref{fig:snap}C reveals that ions (Cl$^-$ in this case) enclose themselves with clusters of water molecules.
To put this into a quantitative perspective, we analyze the local water densities around the ions in bulk water and in the gel phase, see \Fig~\ref{fig:RDFions}A.
The main peak, corresponding to the  first hydration shell, does not weaken that much upon entering from water into the gel.
Instead, the ions preserve their first hydration shell even though the mean water density in PNIPAM is only around one fifth of that in the bulk. We conclude that the ions mostly distribute within the aqueous nanoclusters. 

% cluster definition
Further understanding of the extent of the surrounding water can be obtained by analyzing the hydrating clusters around the ions.
We define a water cluster as the group of the water molecules that are mutually separated by less than 0.35 nm. An ion resides in a cluster if it is separated from any of the water molecules of the cluster by less than 0.4~nm. % (the first hydration shell of I$^-$).
\Figure~\ref{fig:RDFions}B shows the size distributions of all water clusters (dashed line) in the gel as well as ion-hydrating clusters (symbols), which are the clusters formed around ions.
The distribution of water clusters first roughly follows a power-law~\cite{kanduc2018diffusion} $P(N_\trm{w})\sim{N_\trm{w}}^{\!-1.74}$ and crosses over into a roughly exponential decay for larger clusters ($N_\trm{w}>$\,100).
Similarly, the distribution of ion-hydrating clusters also approximately follows an exponential decay for $N_\trm{w}>10$.
The mean ion-hydrating cluster size is $\langle N_\trm w\rangle=\,$120--160, indicated by a green shaded stripe in \Fig~\ref{fig:RDFions}B.
However, the probability of very small ion-hydrating clusters rapidly decays to zero for $N_\trm{w}<10$, indicating that an ion does not get very likely dehydrated.
Another important observation is that the results do not reveal any significant difference between the cations and anions within the accuracy of the data---all the ions behave very similarly.

For the purposes of a simple analysis, we convert the mean ion-hydrating cluster size $\langle N_\trm w\rangle$ into an equivalent size of a spherical water droplet of the bulk water density ($\rho_\trm{w}=32.2$~nm$^{-3}$),
\begin{equation}
\langle N_\trm w\rangle=\frac{4\pi}{3} \rho_\trm{w}{r_\trm{w}}^{\!3}
\end{equation}
which results into the effective droplet size of $r_\trm{w}=$\,0.96--1.06~nm.
%This undermines the continuum picture. 

% Born
A traditional mode of thinking about ionic solvation is based on the continuum picture of the Born free energy~\cite{born1920volumen}, which is particularly powerful for estimating partitioning in homogeneous liquids~\cite{jean1991electrostatic}.
  In this framework, the water and the gel phases are regarded as homogeneous background media with relative permittivities $\varepsilon_\trm{w}$ and $\varepsilon_\trm{g}$, respectively. The transfer free energy is then associated with the change of the Born free energy of an ion with charge $q$ as~\cite{born1920volumen,leunissen2007ion, de2008spontaneous, loche2018breakdown} 
\begin{equation}
\Delta G_\trm{B}=\frac{q^2}{8\pi\varepsilon_0 a_\trm{B}}\left(\frac{1}{\varepsilon_\trm{g}}-\frac{1}{\varepsilon_\trm{w}}\right)
\label{eq:simple}
\end{equation}
Here, $a_\trm{B}$ is the effective Born radius of the ion~\cite{still1990semianalytical, onufriev2002effective}, which characterizes its effective size, usually regarded as the location of the first hydration layer. % in terms of the distance to the molecular surface.
The first hydration layer  can be deduced from water density distributions around the ions ($a_\trm{B}=\,$0.23--0.36~nm, cf.~\Fig~\ref{fig:RDFions}A).
Using evaluated values for the relative permittivities in our model, $\varepsilon_\trm{w}=67$ for bulk water and $\varepsilon_\trm{g}=8.5$ for the gel (see Methods section), we obtain 
$\Delta G_\trm{B}=20$--$30$~kJ/mol. 
But since in our system the ions in the gel are encapsulated by water clusters
(\Fig~\ref{fig:RDFions}C shows a simplified picture), the effective Born radius should rather reflect the size of the ion along with its hydration shell of effective size $r_\trm{w}$.  In other words, upon the transfer of an ion from water into the gel, the dielectric environment around the ion changes only beyond the radial distance of $r=a_\trm{B}$.
Therefore, we can conveniently assume $a_\trm{B}=$\,1~nm in \Eq~\ref{eq:simple}, which  gives us
$\Delta G_\trm{B}=+$7~kJ/mol for both cations and anions.
Note that \Eq~\ref{eq:simple} is symmetric on the sign of the ionic charge $q$ and strictly positive as long as $\varepsilon_\trm{w}>\varepsilon_\trm{g}$.

%Note also that the Born solvation model does not explicitly include the charge asymmetry of the water and polymer molecules, and consequently yields symmetric results for cations and anions.

%This simple continuum picture is clearly not good enough! First, it predicts a total asymmetry between cations and anions of same sizes, since the Born electrostatic interaction scales as $q^2$. Second, it does not explain the negative values of $\Delta G$ for cations. Third, differences of $\Delta G$ between cations and ions are around $100$~kJ/mol in the simulations, whereas the theory gives only a few kJ/mol difference (due to differences in the size $r_0$).

%Potential inside water clusters
%All ions are preferentially solvated inside water clusters, but as we have seen, cations have higher affinity to clusters than anions.

\begin{figure*}[t!]\begin{center}		
\begin{minipage}[b]{0.4\textwidth}\begin{center}
\includegraphics[width=\textwidth]{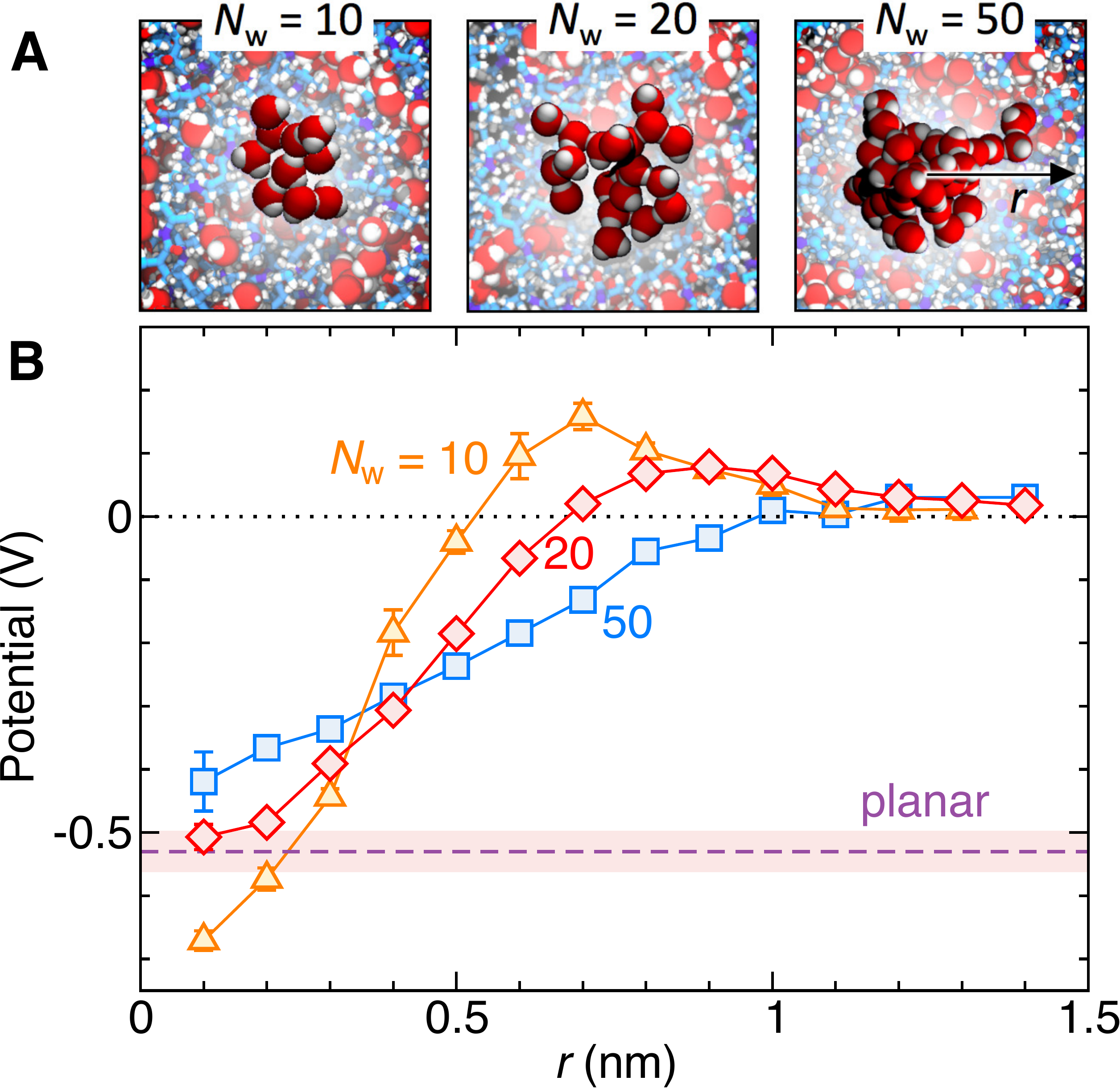}
\end{center}\end{minipage}\hspace{5ex}     
\begin{minipage}[b]{0.39\textwidth}\begin{center}
\includegraphics[width=\textwidth]{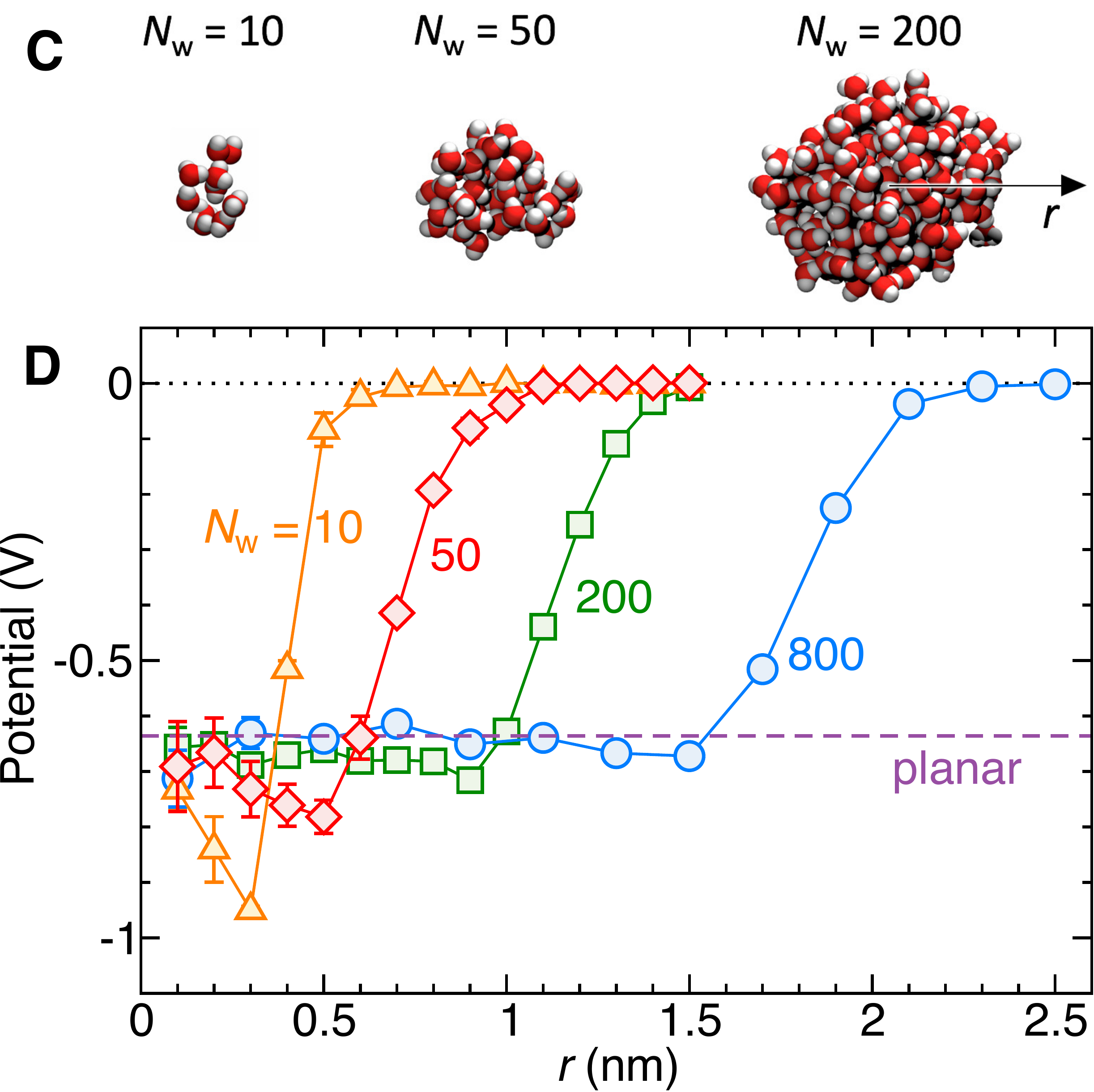}
\end{center}\end{minipage}
\caption{
(A) Snapshots of water clusters of various sizes $N_\trm{w}$ in PNIPAM.
(B)~Radial dependence of the electrostatic potential from the center of a water cluster. 
The horizontal dashed line shows the potential drop across a planar PNIPAM--water interface, $\psi_\trm{s}$, where the thickness of the shaded region represents the numerical uncertainty
(see \SItext).
(C)~Snapshots of water droplets of various sizes $N_\trm{w}$ in vapor.
(D)~Radial dependence of the electrostatic potential from the center of a water droplet. 
The horizontal dashed line shows the potential drop across a planar vapor--water interface.
}
\label{fig:Rpotential}
\end{center}\end{figure*}

% cluster potential
The analysis thus far demonstrated no differences in the structure and distribution of monoatomic ions inside the gel. Since ions are enclosed by a rather bulky water clusters, this raises the question of the influence of these clusters on the containing charge.
An effect that is not accounted for in the implicit Born solvation model is the polarization of water clusters, which comes about due to preferential orientation of water molecules at the cluster interface. 
The polarization at the cluster interface arises from the fact that the water molecules in this region are subject to an anisotropic environment.
%Surface term is there because it arises from the reversible work associated with the ion crossing the air-liquid interface.\cmt{K} 

To answer this question, we evaluate the electrostatic potential inside water clusters of various sizes (\Fig~\ref{fig:Rpotential}A). The radial dependence of the potential from the center of a cluster is plotted in \Fig~\ref{fig:Rpotential}B for three different cluster sizes.
Interestingly, in spite of an irregular, not well-defined shape, the interior of a cluster seems to universally acquire a well-defined potential of $\psi_\trm{cl}\approx-$0.5~V with respect to the surrounding `dry' part of the gel.
In the same plot we also indicate the potential drop of the planar macroscopic PNIPAM--water boundary ($\psi_\trm{s}=-$0.53~V, as obtained from separate simulations, see \SItext).
Already a few water molecules ($N_\trm{w}\sim10$) are enough to generate the potential drop in the center that is almost equivalent to the drop of a macroscopically large interface.
  This means that an ion enclosed in a cluster is directly subjected to the potential drop at the interface. 
Cations gain favorable negative contribution of $e\psi_\trm{cl}=-$49~kJ/mol, whereas anions  are penalized by this same amount, $+$49~kJ/mol.
 
The phenomenon of the interface potential is well known from the context of the water--vapor and water--oil interface~\cite{brodskaya2002molecular, dang2002molecular, znamenskiy2003solvated,hagberg2005solvation, lee2007hydration, hunenberger2011single, beck2013influence, caleman2011atomistic, vacha2011orientation}. 
For the sake of comparison we also show a similar analysis for water nanodroplets in vapor (\Fig~\ref{fig:Rpotential}C, D). In all the droplets, even the smallest one (composed of 10 molecules), the potential reaches around $-0.6$~V with respect to vapor, which is the same as the potential drop at a planar water--vapor interface (horizontal dashed line).
The analysis of potential inside the clusters and droplets show similar behavior and reveals essentially the same physics.
  Note that  the potential profiles tend to smear out for larger clusters because
  the center of a larger cluster may reside outside the cluster owing to its random-walk shape.

\begin{figure}[h]\begin{center}
\begin{minipage}[b]{0.4\textwidth}\begin{center}
\includegraphics[width=\textwidth]{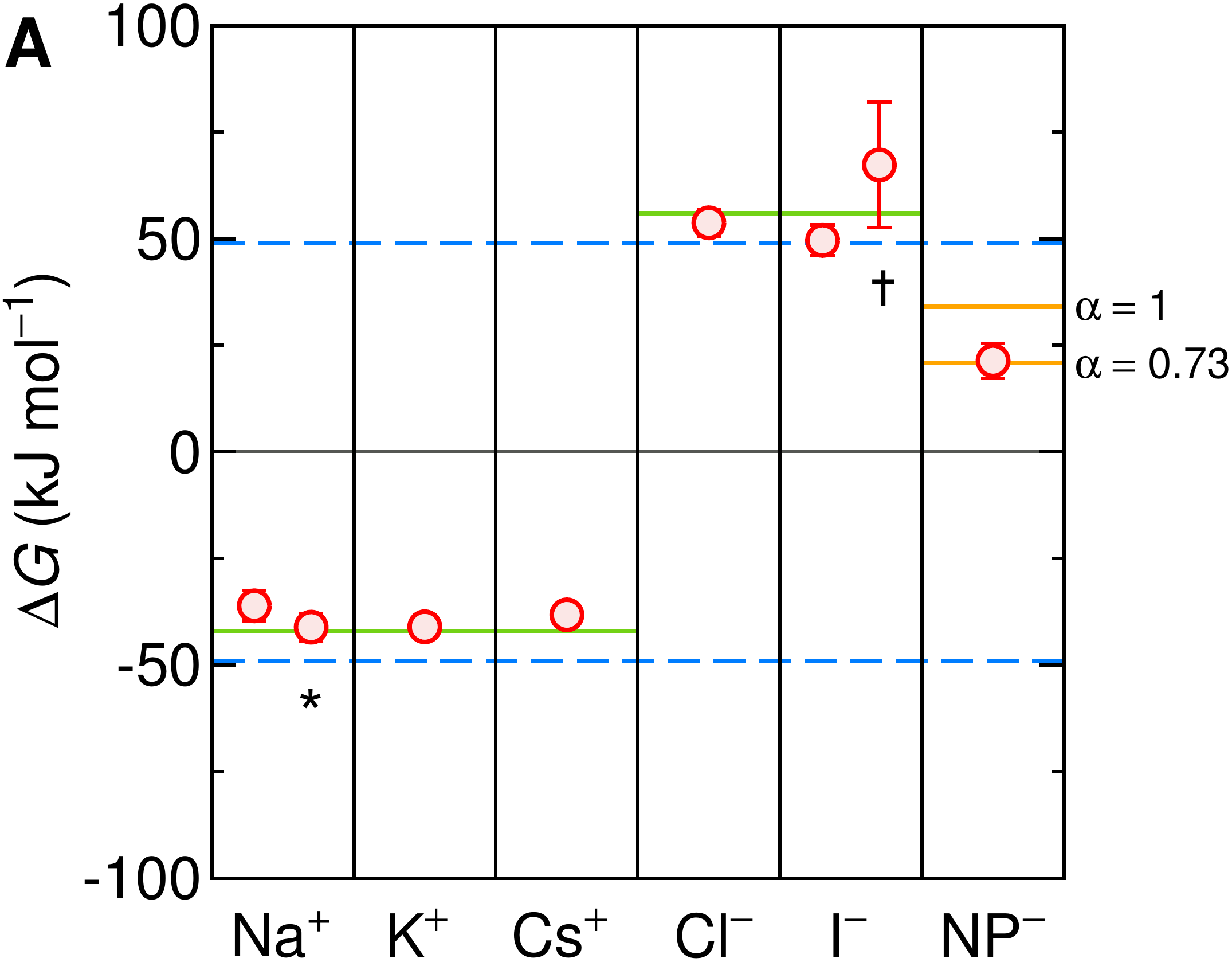}
\end{center}\end{minipage}\vspace{3ex}
\begin{minipage}[b]{0.4\textwidth}\begin{center}
\includegraphics[width=\textwidth]{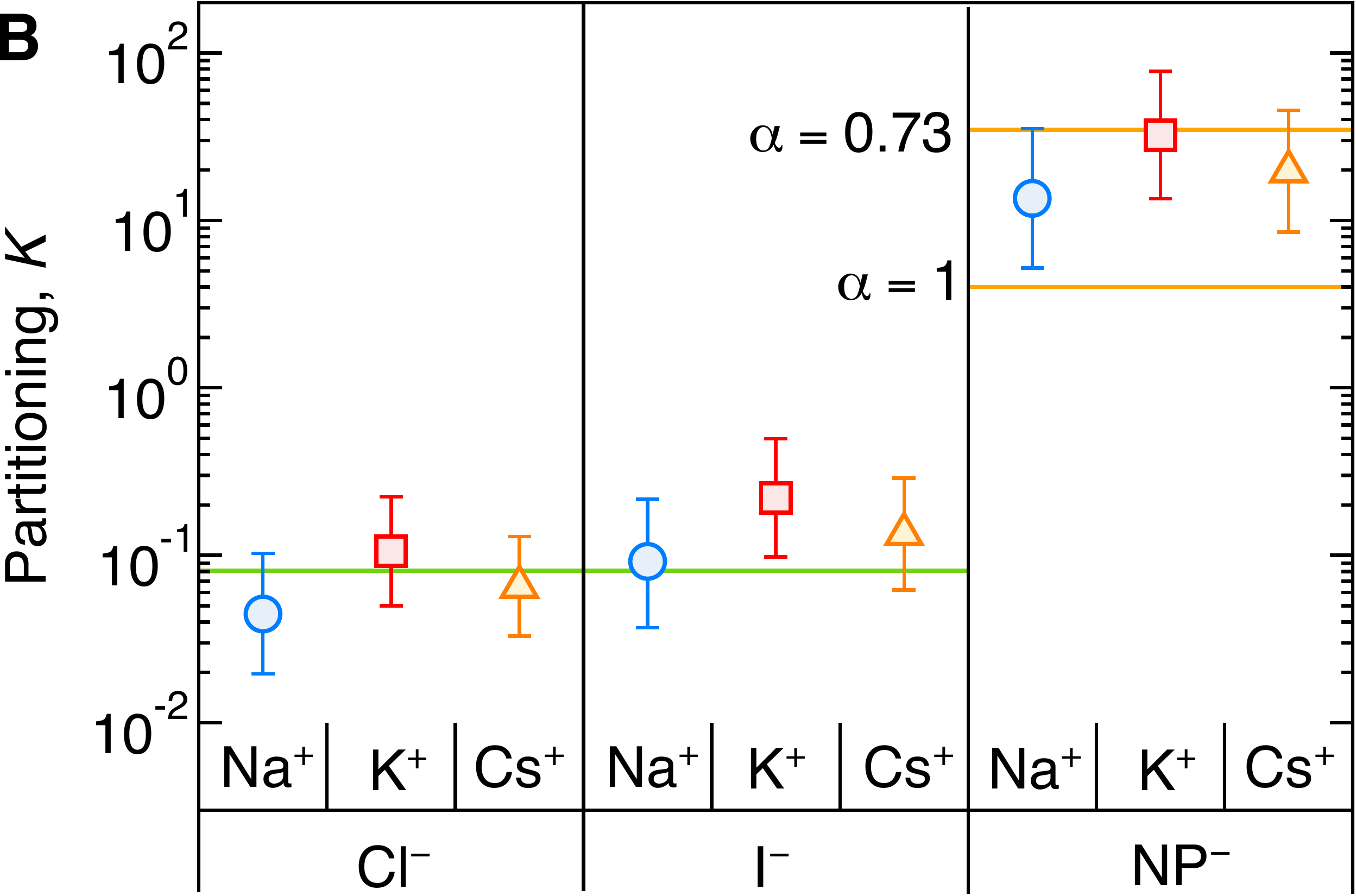}
\end{center}\end{minipage}
\caption{(A) Transfer free energies of ions from simulations (red circles); same as in \Fig~\ref{fig:ions}. The blue dashed lines depict the values 
$\pm e \psi_\trm{cl}$, the green solid lines are the predictions of \Eq~\ref{eq:dGfinal} for monoatomic ions, and the orange lines are the predictions of \Eq~\ref{eq:dGM2} with $\alpha=0.7$ and $1$ for NP$^-$.
(B)~Partition ratios $K$ from simulations computed \via\ \Eq~\ref{eq:K} assuming 1:1 electrolytes for different cation--anion combinations (symbols).
The green line is the prediction of \Eq~\ref{eq:Ksaltapp} and the orange lines are the predictions of \Eq~\ref{eq:KMapprox} with $\alpha=0.73$ and $1$ for combinations involving NP$^-$.
\chgA{
The results are based on the Jorgensen force field for ions~\cite{jorgensenFFNaCl, jorgensenFFI, jensen2006halide}, and where marked additionally $^\star$\AA qvist~\cite{aqvist1990ion} for Na$^+$ and
$^{\dagger}$Dang~\cite{dang1992nonadditive,rajamani2004size} for I$^-$.  
}
}
\label{fig:ions2}
\end{center}\end{figure}

In \Fig~\ref{fig:ions2}A we again plot the transfer free energies from the simulations (red circles) along with the contribution from the cluster potential, $\pm e \psi_\trm{cl}$ (blue dashed lines).
Clearly, the cluster potential alone already roughly captures the difference of simulation data between the monoatomic cations and anions.
The remaining deviations can be reconciled by adding the Born solvation component, {\em viz.}
\begin{equation}
\Delta G^{(i)}= q_i\psi_\trm{cl}+\Delta G_\trm{B}
\label{eq:dGfinal}
\end{equation}
Accepting the Born part of 7~kJ/mol (assuming $a_\trm{B} =\,$1~nm in \Eq~\ref{eq:simple}), the
 predictions of \Eq~\ref{eq:dGfinal} are plotted by solid green lines in \Fig~\ref{fig:ions2}A and capture the data very well.

With the transfer free energy at hand, we are finally in the position to tackle the ion partitioning.  
The {\em individual} partition ratio of ion $i$ is defined as the Boltzmann factor of its transfer free energy,
\begin{equation}
\tilde K^{(i)}=\rme^{-\Delta G^{(i)}/\kB T}
\end{equation}
which does not include collective electrostatic effects from other ions, and should not be confused with the partition ratio $K^{(i)}$ of ion $i$.
Because of the electroneutrality requirements in the gel, the partitionings $K^{(i)}$ of different ion species are coupled and depend on the electrolyte composition (see \SItext).
In the simplest case of a 1:1 electrolyte, the concentrations of cations ($+$) and anions ($-$) are equal and their collective partition ratio is given as the geometric mean of both individual partition ratios
%\begin{equation}
%K^\trm{(salt)}=\exp\left(-\frac{\Delta G^{(+)}+\Delta G^{(-)}}{2\kB T}\right)
%\end{equation}
\begin{equation}
K^\trm{(salt)}=\sqrt{\tilde K^{(+)}\tilde K^{(-)}}
\label{eq:K}
\end{equation}
In \Fig~\ref{fig:ions2}B we plot salt partitioning obtained from the simulation transfer free energy data (symbols) for all combinations of cation and anion species of a 1:1 salt.
For the monoatomic combinations, the partition ratios fall inside the window $K^\trm{(salt)}=0.04$--$0.2$, which is also in the range of reported experimental values for numerous polymers (including PNIPAM) with water content of $K_\trm{w}=0.2$ (\Fig~\ref{fig:K}).

An important observation is also that partitioning does not significantly depend on the type of salt (\ie, ion-specific effects are below the numerical resolution), 
which is in line with simulation studies of core-shell PNIPAM membranes~\cite{adroher2017conformation} as well as with experimental observations~\cite{kawasaki2000partition}.
An easy explanation for this weak ion specificity based on our picture (\Fig~\ref{fig:RDFions}C) is that their specific character is shielded by the surrounding strongly bound hydration shell, that is, the micro-environment does not specifically change upon insertion into the polymer. 
But we can further speculate that the ion specificity may come forth once the hydration cluster start to vanish, which may happen  at even lower hydrations or for larger and more chaotropic ions.

For further discussions it is useful to define the cluster-potential partition ratio
\begin{equation}
K_\trm{cl}=\rme^{-e\psi_\trm{cl}/\kB T}\simeq 3.3\times 10^7
\label{eq:Kcl}
\end{equation}
which represents a hypothetical partition ratio of a cation solely due to the cluster potential, and conversely 
$K_\trm{cl}^{-1}\simeq 3.0\times 10^{-8}$ for an anion. However, these enormous values are not disclosed in the total salt partitioning, since they exactly cancel out ({\em cf.} \Eq~\ref{eq:K}). This is because of equal amounts of cations and anions that are enclosed by the clusters, and therefore the cluster potential remains ``concealed''.
Namely, the theoretical approximation given by \Eq~\ref{eq:dGfinal} together with \Eq~\ref{eq:K} for a monovalent salt leads to
\begin{equation}
K^\trm{(salt)}\simeq\rme^{-\Delta G_\trm{B}/\kB T}
\label{eq:Ksaltapp}
\end{equation}
where only the contribution from the Born solvation survives and yields $K^\trm{(salt)}\simeq$\,0.08. This value, indicated by a green solid line in \Fig~\ref{fig:ions2}B, matches the simulation values for the monoatomic ions considerably well.
 For comparison, in the standard homogeneous Born solvation approach, where one uses the position of the first hydration shell as the effective Born radius, which then yields the Born part of 20--30~kJ/mol (see above), results into substantially too low values $K^\trm{(salt)}\simeq$\,$10^{-5}$--$10^{-3}$.

%Combinations with NP$^-$ as an anion enhances the partitioning, which we will tackle later.

\chgA{
The modified Born solvation model (\Eq~\ref{eq:simple}) furthermore predicts that the ionic solvation should decrease if the ion hydration layer (and with that the Born radius $a_\trm{B}$) decreases, which happens, for instance, when the polymer gets further dehydrated upon heating. Indeed, simulations at higher temperatures (thereby at the equilibrium water amount, $K_\trm{w}$) show that NaCl partitioning goes down, which is quantitatively well captured by the Born solvation model (see \SItext). In the opposite limit of a very swollen gel at low temperatures, the Born model correctly predicts that when} $a_\trm{B}$ tends to infinity and  $\varepsilon_\trm{g}$ tends to $\varepsilon_\trm{w}$, then $\Delta G_\trm{B}$ vanishes and $K^\trm{(salt)}$ approaches unity.

\chgA{Finally, our conclusions for salt partitioning are based on single-ion transfer free energies, valid at low enough concentrations. Certainly, one can expect that at higher salt concentrations, the ion--ion interactions could add an essential contribution to their solvation. We tested the free energy calculations also at finite salt concentrations and concluded that the single-ion results should be valid up to at least several 100~mM of bulk salt concentrations (see \SItext).
}

\subsection*{Molecular ions}

% bulk contribution ∆sG[Igz] is called the intrinsic single-ion solvation free energy, because it exclusively arises from the interaction between the ion and its local solution environment
% Born model represent an estimate for the intrinsic (rather than real) solvation free energy of the ion.

Quite generally, large molecular ions (\eg, ionized medicinal molecules, some pharmacological molecules) that possess a considerable electroneutral part can behave very differently from small monoatomic ions~\cite{palasis1992permeability, guilherme2003hydrogels, molinaPolymer2012}.
Also, the physics involved in their hydration is expected to be comparatively more complex. Their geometry and charge density are not spherically symmetric.
Our representative for a molecular ion is NP$^-$, where the deprotonation of the OH group for pH\,>\,7.15 creates an ionized oxygen center~\cite{nitrophenol-pKa}. 
As we have already seen, the transfer free energy $\Delta G$ clearly deviates from the trends of the monoatomic ions (see \Fig~\ref{fig:ions2}A).

\begin{figure}[t]\begin{center}
\begin{minipage}[b]{0.35\textwidth}\begin{center}
\includegraphics[width=\textwidth]{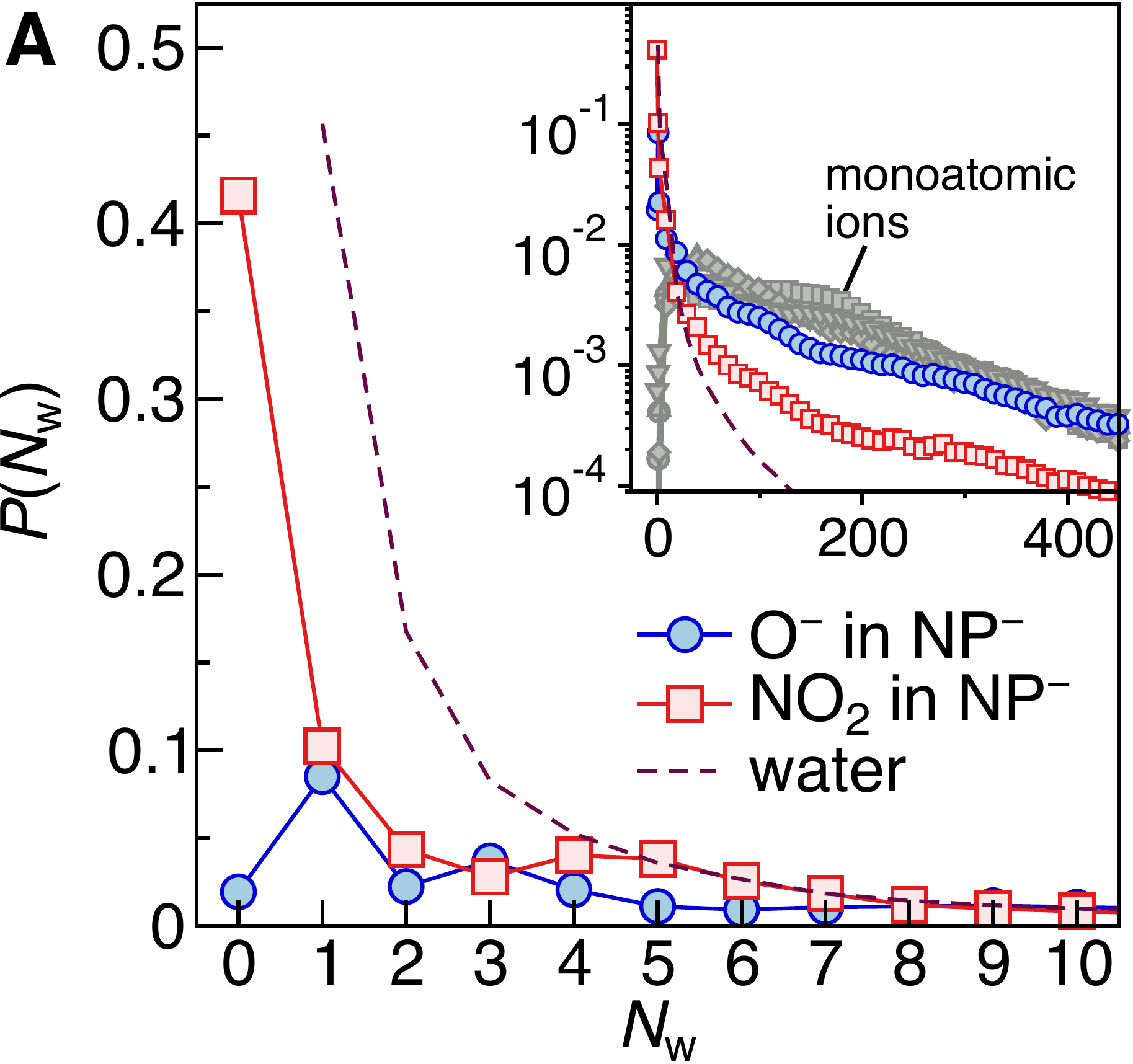}
\end{center}\end{minipage}\vspace{2ex}
\begin{minipage}[b]{0.24\textwidth}\begin{center}
\includegraphics[width=\textwidth]{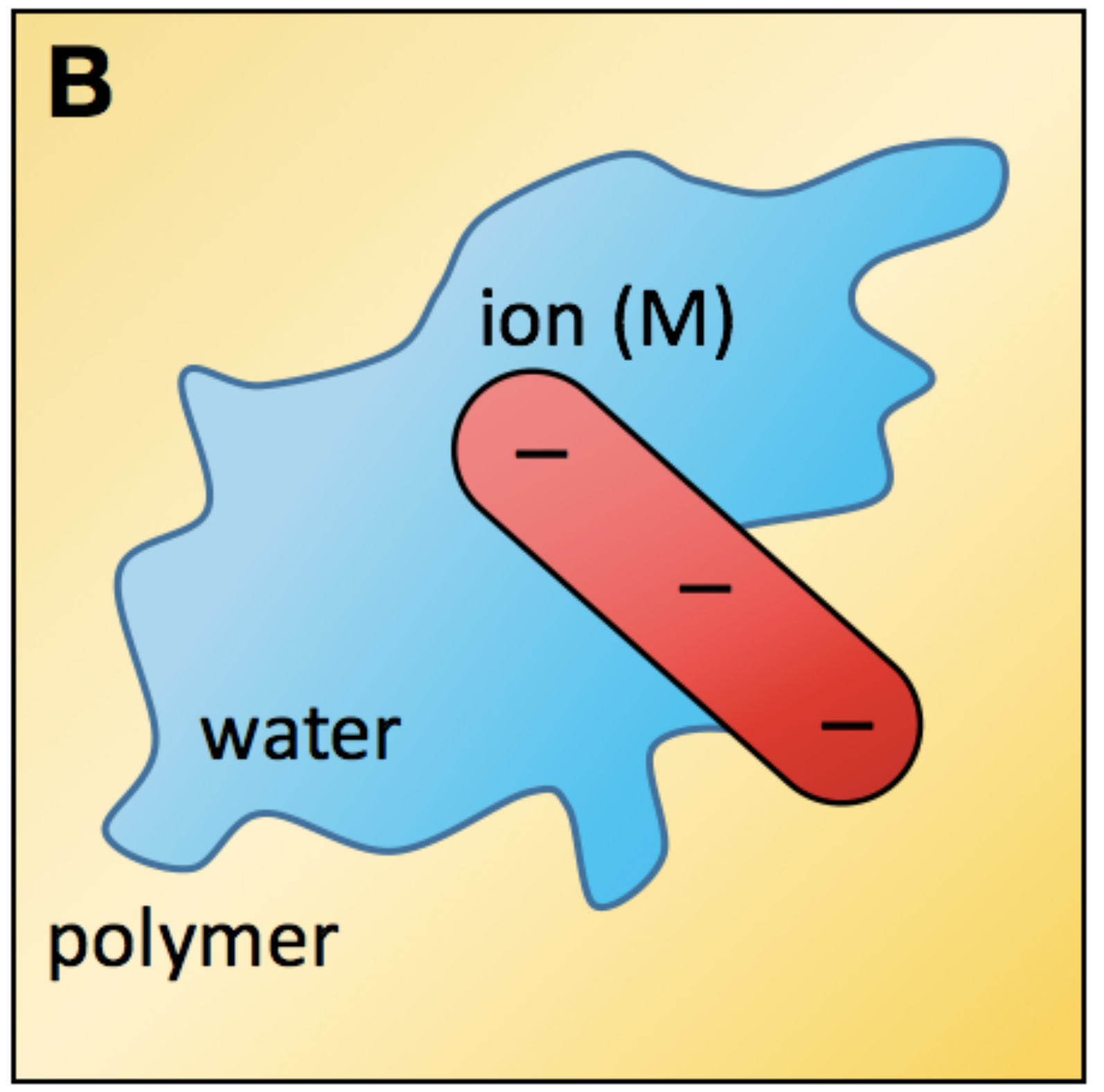} %\phantom{X}
\end{center}\end{minipage}
\caption{
(A) Ion-hydrating cluster size distribution  $P(N_\trm{w})$ around the oxygen atoms in NP$^-$ in comparison with monoatomic ions (shown by gray symbols in log--linear plot in the inset).
The dashed line shows the size distribution of all water clusters in the gel.
(B)~Schematic continuum representation of a large molecular ion that partially sticks out of the hydrating cluster (partial dehydration).}
\label{fig:NPdep}
\end{center}\end{figure}

%Furthermore, the charge distribution at the de-ionized center can be distributed among several neighboring atoms. In our example of a molecular ion, NP$^-$, the electronic density around the deprotonated oxygen (O$^-$) is delocalized across the aromatic ring into the nitro (NO$_2$) group (see \Fig~\ref{fig:NPdep}A). As a consequence, the O$^-$ possesses a partial charge of only $q_1=-0.73~e$, whereas the NO$_2$ group gains the rest $q_2=-0.3~e$. This redistribution of the  ionic charge across the molecule can have important consequences on the partitioning, which we analyze in the following.
%Nevertheless, in a very simplified picture we can consider the NP$^-$ ion as having the O$^-$ atom totally hydrated, whereas NO$_2$ completely dehydrated, as schematically shown in \ref{fig:NPdep}C. We are dealing with an asymmetry where one part of the ion (O$^-$) is subjected to the cluster potential, whereas the other part (NO$_2$) is not.

We first take a look at the hydration of the oxygen atoms in NP$^-$: the deprotonated one O$^-$ and the two in the nitro NO$_2$ group, see \Figure~\ref{fig:NPdep}A. For comparison, the hydration of monoatomic ions (similar as in \Fig~\ref{fig:RDFions}) is shown by gray symbols.
It can be seen that O$^-$ is not hydrated to that extent as the monovalent ions. For instance, the probability that the hydrating cluster consists of more than 5 water molecules \chgA{(by summing up the probabilities $P(N_\trm{w})$ for $N_\trm{w}\ge 6$)} is only around 0.8, whereas it approaches 1 in cases of the monoatomic ions.
On the other hand, the NO$_2$ group is only weakly hydrated (with 0.66 probability of the hydration shell smaller than 5 water molecules).
The incomplete hydration of NP$^-$ can be attributed to a
lowered charge density at the O$^-$ site, which exerts a more moderate electric field on the surrounding water molecules, resulting in a more labile and shorter-ranged hydration structure.
This is because the lone pair from the oxygen  delocalizes \via\ conjugation to the benzene ring and nitro group, which in turn makes the NO$_2$ group slightly charged (see \SItext\ for distribution of charges).

In a simple mechanistic picture that transpires from the discussion above, the charged O$^-$ and NO$_2$ centers of NP$^-$ manage to occasionally escape the water cluster, as schematically depicted in \Fig~\ref{fig:NPdep}B. With this in mind, we construct a phenomenological description for the free energy of a molecular ion~(M) as
\begin{equation}
\Delta G^\trm{(M)}\simeq \Delta G_\trm{neut}^\trm{(M)}+\alpha q\psi_\trm{cl}+\Delta G_\trm{B}
\label{eq:dGM2}
\end{equation}
which is composed of the contribution of the neutral, non-ionized form of the molecule (first term), the modified cluster contribution (second term), and the Born contribution (last term). The latter is for simplicity assumed to be the same as for monoatomic ions.
In our case, the first term is the transfer energy of nitrophenol, NP$^0$ (non-ionized NP$^-$), which is $\Delta G_\trm{neut}^\trm{(M)}=-$22(1) kJ/mol as obtained previously~\cite{kanduc2019free}. As has been shown, the transfer free energy of a neutral molecule scales to very good extent with the molecular surface area, $\Delta G_\trm{neut}^\trm{(M)}\propto A_\trm{m}^\trm{(M)}$~\cite{kanduc2019free}.
The second term is the cluster-potential contribution, now multiplied by a {\em hydration} parameter $\alpha$, a phenomenological parameter that accounts for incomplete hydration. 
If the entire charged part lies inside the cluster, then  $\alpha=1$ as for monoatomic ions,  whereas $\alpha<1$ represents situations where the charge is partially dehydrated and partially evading  the influence of the cluster potential, which decreases the free energy of a negative charge.
The fit of \Eq~\ref{eq:dGM2} to the simulation data point of NP$^-$ in \Fig~\ref{fig:ions2}A gives $\alpha=0.73$. 
For comparison we show also the prediction for $\alpha=1$, which yields a too high value.

Moving on to the partitioning, we show in \Fig~\ref{fig:ions2}B the partition ratios of the 1:1 salts with NP$^-$ as the anionic component.
The nitrophenol salts partition in the gel much more than monoatomic salts do.
This can now be easily understood by using the approximate expressions for a monoatomic cation $\Delta G^{(+)}$ and for the molecular ion $\Delta G^\trm{(M)}$ (\Eqs~\ref{eq:dGfinal} and~\ref{eq:dGM2}, respectively) in \Eq~\ref{eq:K}, which results in the expression
\begin{equation}
K^\trm{(M)}\simeq\sqrt{K_\trm{neut}^\trm{(M)}\,K_\trm{cl}^{1-\alpha}}\,K^\trm{(salt)}
\label{eq:KMapprox}
\end{equation}
where $K_\trm{neut}^\trm{(M)}=\exp(-\Delta G_\trm{neut}^\trm{(M)}/\kB T)$ is the partition ratio of the non-ionized form of the molecule
 and $K^\trm{(salt)}=0.08$ is the theoretical prediction for the partitioning of monoatomic salts.  Notably, because of an incomplete hydration ($\alpha<1$) of one of the salt components (NP$^-$ in our case), the cluster potential (expressed as $K_\trm{cl}$, \Eq~\ref{eq:Kcl}) influences  the partitioning explicitly,  in contrast to  the case of monoatomic ions.
The predictions of \Eq~\ref{eq:KMapprox} for $\alpha=0.73$ and $\alpha=1$ are depicted in \Fig~\ref{fig:ions2}B by orange lines and demonstrate that the cluster potential has an important contribution to the partitioning.

Moreover, in many practical scenarios, larger molecular ions are not a component of the electrolyte, but are instead typically introduced in trace (submillimolar) amounts into a system containing a simple, monoatomic monovalent salt (\eg, NaCl). Such are typical cases of charged drugs/pharmaceuticals in a physiological solution or reactants in catalytic experiments.
In the case of a 1:1 salt of a much larger concentration $c_0$ than the concentration $c_\trm{M}$ of the molecular ions, %(typically acompanying with changes in OH$^-$ and H$_3$O$^+$ concentraions in submillimolar domain)
 the expression for the partitioning of the molecular ions reads (see \SItext)
\begin{equation}
K^\trm{(M)}= \tilde K^\trm{(M)}\sqrt{\frac{\tilde K^{(+)}}{\tilde K^{(-)}}}
\label{eq:KM11}
\end{equation}
For our system, we obtain the partitioning of NP$^-$ ions in excess of NaCl in the hydrogel $K^\trm{(M)}=4 \times 10^3$, which is an enormous value compared to the partitioning of NaCl salt! A full calculation for the cases with two anions of arbitrary concentrations shows that the drastic consequence of the large partitioning of species M is that only relatively small amounts of it ($\sim$~mM) are needed to almost completely exchange the smaller anion, \ie, the partitioning of the latter decreases by orders of magnitudes (see \Fig~S3 in the \SItext). 

Furthermore, the non-ionized form of the molecule, NP$^0$, partitions with a very similar ratio, $K_\trm{neut}^\trm{(M)}=3\times 10^3$~\cite{kanduc2019free}. This may seem counterintuitive and against a common knowledge, which would anticipate that charged molecules should be significantly less attracted to a weakly hydrated and hydrophobic gels than their neutral counterparts. 
In order to understand this non-apparent and surprising outcome, we use our phenomenological theoretical framework, which helps us to explain the main qualitative trends.
Inserting the theoretical expressions (\Eqs~\ref{eq:dGfinal} and~\ref{eq:dGM2}) into \Eq~\ref{eq:KM11}, gives us
\begin{equation}
K^\trm{(M)}\simeq K^\trm{(M)}_\trm{neut}K_\trm{cl}^{1-\alpha} K^\trm{(salt)} 
\label{eq:KM11app}
\end{equation}
In words, the partitioning of a charged molecule in excess of salt is proportional to the partitioning of its non-ionized form, $K^\trm{(M)}_\trm{neut}$, which is then modified by the factor $K_\trm{cl}^{1-\alpha} K^\trm{(salt)}$ due to ionization. In case the charged molecular centers are fully hydrated ($\alpha=1$), the ionization decreases the partitioning only by the factor of $K^\trm{(salt)}\simeq 0.08$. On the other hand, if the charged centers get partially dehydrated ($\alpha<1$), this blows up the partitioning in an exponential manner\chgA{---in our case with $\alpha=0.73$, this adds a factor of $K_\trm{cl}^{1-\alpha}\simeq 10^2$.}
One needs to be aware that \Eq~\ref{eq:KM11app} is very rough and cannot yield reliable quantitative results, but it nevertheless gains important qualitative insights into the partitioning  governed by water clusters due to incomplete hydration.
% can have enormous influence and the exact value is extremely sensitive to the details.

Our simulations as well as the theoretical analysis show that large ionized molecules can exhibit enormous partitioning ($K\sim 10^3$) in weakly hydrated gels, which was reported in numerous experiments of charged pharmaceuticals in PNIPAM hydrogels~\cite{palasis1992permeability, guilherme2003hydrogels, molinaPolymer2012}.
Notably, the charge does not necessarily impede the partitioning but can also enhance it.

\section*{Conclusion}

The understanding of the solvation and structure  of charged molecules within dense polymer systems remains rudimentary owing to a high complexity of the molecular mechanisms involved. 
Resorting to a classical atomistic simulation framework, we investigate partitioning of small monoatomic and larger molecular ions in dense, weakly hydrated neutral polymers. 
The simulations reveal that ions are enclosed by hydrating, nanosized water clusters that accommodate them within a principally hydrophobic surrounding.

% monoatomic ions
The estimates for partitioning of monoatomic salts between 0.04--0.2 from simulations agree well with experiments for a wide range of polymers of a similar water content. This partitioning is much larger than inferred from a simple Born solvation model with an effective homogeneous dielectric constant. 
The hydration of monoatomic cations and anions is structurally almost indistinguishable. However,  the polarization stemming from the water-cluster interfaces induces a potential drop of around $-$0.5~V and renders a substantial difference in the transfer free energies of ions from water into the gel: negative for cations and positive for anions.
Nevertheless, the explicit impact of the cluster potential on the thermodynamic partitioning disappears for monoatomic salts because of overall electroneutral sets of ions that are subjected to this potential. This fact makes the modified Born solvation model that accounts for a one-nanometer-thick hydration shell sufficient. 

% molecular ions
The story becomes more intricate for larger molecular ions, in our study represented by 4-nitrophenolate. Because of its labile hydration shell, the charged parts of the molecule can occasionally ``step out'' of the cluster, thereby escaping the influence of its electrostatic potential.
This has far-reaching consequences on the thermodynamics. The cluster potential is no longer completely compensated, and it thus reveals its direct influence on the total partitioning of the ions, which is by itself an interesting phenomenon. 
% significance
Namely, the cluster potential is a special type of an interface potential (the water--air interface potential being the most prominent example), and as such it may be expected to be thermodynamically inaccessible and nullified in the overall salt partitioning due to the net charge in action. However, the fragmented water clusters together with the asymmetry in hydration strength of ion species involved, elude this paradigm.
%The cluster potential is a special type of an interface potential (the water--air interface potential being the most prominent example). For macroscopic systems, the interface potential is not influential due to the net charge in action, and is therefore thermodynamically inaccessible.
%However, in the systems of dense and weakly hydrated polymers, the clustered water enables ...and the interface potential has a direct influence on the thermodynamics.
% influence
A direct consequence of the incomplete cancellation of the cluster potential is a higher ionic partitioning. In fact, ionizing the molecule may even enhance its solubility in the gel, which opposes the standard picture of ion solvation.
\chgA{Indeed, high partitioning has been observed in experiments with charged organic molecules~\cite{guilherme2003hydrogels, molinaPolymer2012}. Also, the high partitioning of 4-nitrophenolate explains its fast reduction kinetics in collapsed PNIPAM-based nanoreactors~\cite{dzubiellaAngew2012}.}

Our study provides important and very fundamental insights into the microscopic mechanisms behind the solvation of ions and charged molecules not only in dense hydrogels but in other amorphous matter with water micro-phases as well. The notion of the surface potential is extremely vital for various research direction, ranging from desalination membranes to biomedical applications and pharmacokinetics.

% for instance the prediction of the transport and partition coefficients of ionized drugs are of key requirement for the pharmaceutiical industry.

%Surface surface term, because it arises from the reversible work associated with the ion crossing the air-liquid interface, 
%The interface potential is experimentally elusive, because it is thermodynamically inaccessible~\cite{hunenberger2011single} and requires extra-thermodynamic assumptions.

\section*{Methods}
%\footnotesize
\subsection*{Model and force fields}

The atomistic computer model of the collapsed PNIPAM polymers with sorbed water is adopted from our previous works~\cite{kanduc2018diffusion, kanduc2019free}.
In this model, 48 atactic PNIPAM chains built out of 20 monomers are condensed together with 1325 water molecules (corresponding to the water amount in equilibrium with bulk water) in a cubic simulation box (of size $\sim$6~nm) at 340~K and isotropic pressure of 1~bar. 

 For PNIPAM polymers, we use the recently improved OPLS-based model by Palivec {\em{et al.}}~\cite{palivecheyda2018} with an {\em ad hoc} parameterization of partial charges that reproduces the thermo-responsive properties better than the standard OPLS-AA force field.
For water we use  the  SPC/E model~\cite{spce}.
We use the force field by Jorgensen and coworkers~\cite{jorgensenFFNaCl, jorgensenFFI, jensen2006halide} for all monoatomic ions, 
and additionally by \AA qvist~\cite{aqvist1990ion} for Na$^+$ and Dang~\cite{dang1992nonadditive,rajamani2004size} for I$^-$.
For nitrophenolate ion NP$^-$, we use the OPLS-based force field with the  partial charge parametrization ``OPLS/QM1'' from Ref.~\citen{kanduc2017selective}.

\subsection*{Simulations}
All atomistic MD simulations are performed using the GROMACS simulation package~\cite{gromacs, abraham2015gromacs}.
  Electrostatics is treated using the particle-mesh-Ewald method~\cite{PME1,PME2} with a 1 nm real-space cutoff. The Lennard-Jones (LJ) interactions are truncated at
1 nm. The simulations are performed with an integration time step of 2 fs in the constant-pressure (NPT) ensemble with periodic boundary conditions. Temperature was maintained at 340 K, using the velocity-rescale thermostat~\cite{v-rescale} with the time constant of 0.1 ps. The isotropic pressure was maintained at 1~bar using the Berendsen barostat~\cite{berendsenT} with the time constant of 1~ps.

\subsection*{Free energy calculations}
\label{sec:MethodsSolvation}
The solvation free energies of ions are calculated using Thermodynamic Integration~(TI)~\cite{frenkel1984new}.
To reduce mutual ion--ion effects (note that the Bjerrum length is 6 nm), we restricted the number of ions in the simulation box to three. The net charge is compensated by applying a uniform neutralizing background charge. 

We first insert three ions of the same type at random positions into an equilibrated polymer system and equilibrate it further with  fully interacting ions for at least 100~ns.
The necessary equilibration time is estimated based on the crossover time of the ion to reach the normal diffusion~\cite{kanduc2018diffusion}. 

During the TI simulations, the Coulombic and LJ interactions of the equilibrated ions are gradually switched off. Introducing a coupling parameter $\lambda\in[0,1]$ that continuously switches the interactions in the Hamiltonian $U(\lambda)$ between the original interactions (for $\lambda$\,$=$\,1) and a non-interacting ion (for $\lambda$\,$=$\,0), the solvation free energy is computed as~\cite{frenkel1984new}
\begin{equation}
G^\trm{TI}=\int_0^1\left\langle\frac{\partial U(\lambda)}{\partial\lambda}\right\rangle_\lambda \rmd\lambda
\label{eq:TI}
\end{equation}
The integration is performed in two stages: We first linearly scale down the charge, while keeping the LJ interactions intact. In the second stage, we scale down the LJ interactions using the ``soft-core'' LJ potentials as implemented in GROMACS in order to avoid singularity problems when the interactions are about to vanish ($\lambda\to0$)~\cite{beutler1994avoiding}. 
The entire TI procedure is composed of 24 individual simulation windows with equidistant $\lambda$ values for the Coulomb part and likewise 24 windows for the LJ part.
The simulation time of each individual $\lambda$ window is 4~ns where the first 3~ns are discarded from the sampling to allow an equilibration of the ion. 
%The estimation of the equilibration time in each individual simulation is based on the drift of the $(\partial U/\partial\lambda)$ output.
All the TI calculations are performed with 5--6 independently equilibrated systems for the Coulomb part and 1--2 systems for the LJ part (note that the LJ part converges considerably faster 
than the Coulomb part, allowing also a much more accurate evaluation). 
The final results are averaged over all three ions in the simulation box and over all the systems.
Even though each ion samples only a local phase space during a short TI time window, the used procedure assures adequately sampled values of the free energy.

% Horinek:
Because of the Ewald summation with periodic boundary conditions, the free energy of solvation depends on the size of the simulation box, and we have to apply the finite-size correction~\cite{hummer1997ion},
\begin{equation}
\Delta G^\trm{corr}_k= \frac{2.837\, q^2}{8\pi\varepsilon_k\varepsilon_0 L}
\label{eq:Fqcorr}
\end{equation}
where $q$ is the ion charge, $\varepsilon_k$ the relative permittivity of medium $k$ [water~(w) or gel~(g)], and $L$ the size of the simulation box.
The correction in our case accounts for around 0.5 kJ/mol in water and 3.8 kJ/mol in the gel, which is comparable to the uncertainties of the evaluated free energies. Therefore, higher-order corrections  to \Eq~\ref{eq:Fqcorr} (\eg, orientational polarization of the solvent due to
periodic images of the ion~\cite{lee2007hydration}) are not necessary.
Finally, the solvation free energies for transferring an ion from vacuum into the medium
are obtained as a difference of the values obtained from TI (corrected for \Eq~\ref{eq:Fqcorr}), 
\begin{equation}
G_{k}=G^\trm{TI}_{k}-G^\trm{TI}_\trm{vac}+\Delta G^\trm{corr}_{k}-\Delta G^\trm{corr}_\trm{vac}
\end{equation}
 Note that $G^\trm{TI}_\trm{vac}$ may be nonzero for molecules (NP$^-$) due to intra-molecular Coulombic and LJ contributions.
 
The evaluated free energies represent the intrinsic single-ion solvation free energies, arising  exclusively from the interaction between the ion and its local solution environment and takes neither the surface potential $\psi_\trm{s}$~\cite{lee2007hydration} nor the compression free energy~\cite{hummer1996free} into account.
We also verified that performing TI on three ions of same kind with a uniform background charge yields same results as performing TI on ion pairs (see \SItext).

\subsection*{The need for free energy calculations}
\label{sec:MethodsDirectK}
 In principle, an alternative simulation setup containing a polymer membrane in water provides a possibility for a direct evaluation of $K$ from the equilibrated ion concentrations inside and outside the membrane.
The minimal width of both the water and polymer slabs would need to be at least $d=$\,3~nm in order to overcome the interface regions of 1~nm (Debye length at 100~mM).
An estimated time for an ion to diffuse from one end of the membrane slab to the other is $d^2/D_\trm{g}$, where $D_\trm{g}\sim$\,10$^{-4}$~nm$^2$/ns~\cite{kanduc2018diffusion} is the diffusion coefficient in the gel membrane. Furthermore, the statistics is hindered by the low ion partitioning $K$, such that the  necessary simulation time for a meaningful sampling would be $t_\trm{sim}\gg d^2/(K D_\trm{g})\sim 10^6$~ns. This is three orders of magnitude longer than for the free-energy calculations and hence far beyond the current rational capabilities.
The direct method lends itself, for example, for higher polymer hydrations~\cite{adroher2017conformation} and implicit-solvent models~\cite{kim2017cosolute}.

\subsection*{Relative permittivity}
The static relative permittivities of bulk water and PNIPAM gel are calculated from the fluctuations of the total dipole moment $\Av{M}$ of the system as~\cite{dommert2008comparative}
\begin{equation}
\varepsilon=1+\frac{1}{3V\varepsilon_0 \kB T} \left(\langle \Av{M}^2\rangle-\langle \Av{M}\rangle^2\right)
\end{equation}
where $V$ is the volume of the simulation box.
The fluctuations are evaluated from the systems without ions.

\subsection*{Electrostatic potential}
In the case of water droplets and clusters, we compute the electrostatic potential along the radial distance from the center of mass (COM) of each target droplet or cluster as
\begin{equation}
\phi(\Av r)=\sum_i\frac{q_i}{4\pi\varepsilon_0|\Av r-\Av r_i|}
\end{equation}
where the summation runs over all the partial charges $q_i$ of the system, located at  positions $\Av r_i$.
For each radial distance $r=|\Av r|$ from the COM, we average the results over the entire solid angle around the droplet or the cluster.
In the case of clusters, we also average the results over all the clusters of a target size $N_\trm{w}$ in each stored simulation frame.

%    we identify the center of mass of each cluster of target size and evaluate the potential along the radial distance r from the COM.
    
The potential across planar water--vapor and PNIPAM--water interfaces are calculated from independent simulations that consist of two separated phases by double integration of the Poisson equation. For a detailed analysis see \SItext.

%This is not yet the total free energy (look further), but only a ``partial'' or ``local'' or ``microscopic'' free energy, because it does not include the surface potential (e.g., of the PNIPAM--water or water--vapor interface).

\normalsize
\subsection*{Conflict of Interest} The authors declare no competing
financial interest.

\subsection*{Associated Content}
The \SItext\ is available free of charge and contains information on:
Partial charges of nitrophenolate;
Free energy calculations at finite salt concentrations;
Temperature dependence;
Potential across a planar interface;
Partitioning of ions;
Ion distribution across the interface

\subsection*{Acknowledgments}
%The authors thank Yan Lu and Matthias Ballauff for useful discussions.
 This project has received funding from the European Research Council (ERC) under the European Union's Horizon 2020 research and innovation programme (Grant Agreement no.\ 646659-NANOREACTOR). M.K.~acknowledges the financial support from the Slovenian Research Agency (research core funding no.\ P1-0055).
W.K.K.\ acknowledges funding from the Deutsche Forschungsgemeinschaft (DFG) \via\ grant NE 810/11.
The simulations were performed with resources provided by the North-German Supercomputing Alliance~(HLRN).

\footnotesize
\setlength{\bibsep}{0pt}
\bibliography{literature}

\end{document}